%% file: submission.tex
\documentclass[lettersize,journal]{IEEEtran}

\usepackage{tikz}
\usepackage{amsmath}
\usepackage{amssymb}
\usepackage{pifont}
\usepackage{multirow}
\usepackage{svg}
\usepackage{subfloat}
\usepackage{placeins}
\usepackage{subcaption} 
\usepackage{soul}

\usepackage{filecontents}

\usepackage{xurl}
\usepackage{cite}

\usepackage{multirow, makecell}
\usepackage{booktabs}

\usepackage{glossaries}

\newacronym{enf}{ENF}{electric network frequency}
\newacronym{snr}{SNR}{signal-to-noise ratio}
\newacronym{rt}{RT}{reverberation time}
\newacronym{dl}{DL}{deep learning}
\newacronym{cnn}{CNN}{convolutional neural network}
\newacronym{crnn}{CRNN}{convolutional recurrent neural network}
\newacronym{rnn}{RNN}{recurrent neural network}
\newacronym{nn}{NN}{neural network}
\newacronym{seq2seq}{seq2seq}{sequence-to-sequence}
\newacronym{air}{AIR}{acoustic impulse response}
\newacronym{pp}{pp}{percentage points}
\newacronym{mfcc}{MFCC}{Mel frequency cepstral coefficient}
\newacronym{dnn}{DNN}{Deep Neural Networks}
\newacronym{svm}{SVM}{Support Vector Machine}
\newacronym{gmm}{GMM}{Gaussian Mixture Model}
\newacronym{roc}{ROC}{receiver operating characteristic}
\newacronym{rmse}{RMSE}{root mean squared error}
\newacronym{tpr}{TPR}{true positive rate}
\newacronym{fpr}{FPR}{false positive rate}
\newacronym{auc}{AUC}{area under the curve}

\usepackage{hyperref}
\usepackage{xspace}
\makeatletter
\DeclareRobustCommand\onedot{\futurelet\@let@token\@onedot}
\def\@onedot{\ifx\@let@token.\else.\null\fi\xspace}
\def\eg{\emph{e.g}\onedot} 
\def\ie{\emph{i.e}\onedot} 
\def\cf{\emph{cf}\onedot} 
 
\def\wrt{w.r.t\onedot} 
\def\etal{\emph{et al}\onedot}
\makeatother

\newcommand{\edited}[1]{\textcolor{black}{#1}}
\newcommand{\minorRev}[1]{\textcolor{black}{#1}}

\begin{document}

\title{EnvId: A Metric Learning Approach for Forensic Few-Shot Identification of Unseen Environments}
\author{Denise Moussa\textsuperscript{*}, Germans Hirsch\textsuperscript{*}, Christian Riess \\
\textit{Friedrich-Alexander-Universität Erlangen-Nürnberg} \\
\{denise.moussa, christian.riess\}@fau.de 

\thanks{* Authors contributed equally to this work.}

}

\maketitle

\begin{abstract}
Audio recordings may provide important evidence in criminal investigations.
One such case is the forensic association of a recorded audio to its recording location. 
For example, a voice message may be the only investigative
cue to narrow down the candidate sites for a crime.
Up to now, several works provide supervised classification tools for closed-set recording environment identification under relatively clean recording conditions.
However, in forensic investigations, the candidate locations are case-specific.
Thus, supervised learning techniques are not applicable without retraining a classifier on a sufficient amount of training samples for each case and respective candidate set. 
In addition, a forensic tool has to deal with audio material from uncontrolled sources with variable properties and quality.
In this work, we therefore attempt a major step towards practical forensic application scenarios. 
We propose a representation learning framework called EnvId, short for environment identification.
EnvId avoids case-specific retraining by modeling the task as a few-shot classification problem.
We demonstrate that EnvId can handle forensically challenging material. 
It provides good quality predictions even under unseen signal degradations, out-of-distribution  reverberation characteristics or recording position mismatches.\footnote{code: \url{https://faui1-gitlab.cs.fau.de/mmsec/few-shot-recording-environment-identification}}

\end{abstract}

\begin{IEEEkeywords}
audio forensics, representation learning, environment identification.
\end{IEEEkeywords}

\section{\edited{Introduction}}
\label{sec:intro}

Inferring the recording location from some audio material can be a highly relevant audio forensic task during criminal investigations.
The analysis of audio material can directly contribute to evidence or at least provide investigative cues for the  reconstruction of crime scenes and the course of events in a case~\cite{zjalic2020digital}.

A reverberant environment, which we also refer to as location, may leave characteristic reverberation traces in audio recordings.
Reverberation, most generally described, results from reflections of sound
waves in an environment. 
For example, when a recording is made of a sound that originates next to a wall (consider, \eg, the breaking of a glass), then the microphone records the primary sound that travels directly from the origin to the microphone, and a reflection of that sound from the wall.
In principle, any obstacle in the environment that interferes with sound waves, \ie, in terms of absorption, transmission and reflection properties, contributes to its reverberation signature.
A simple computational model for reverberation is the convolution of a
signal with an \gls{air}.

Consequently, to identify a certain location, \eg, a specific room in an apartment, one needs to invert the mapping of recorded reverberation traces to the underlying geometry and overall composition of the respective environment.
This mapping is not unique, but from a practical point of view it is oftentimes sufficiently distinctive to support or reject hypotheses that a recording has been made in one out of a set of specific locations.
\edited{Note that this task is different from acoustic scene classification, where the goal is to differentiate between certain types of environments like \emph{public street} or \emph{train station} that share similar background noise profiles~\cite{geiger2013large, 10446177, apostolidis2024visual}.}

So far, several analytic and \gls{dl} methods have been proposed that exploit reverberation cues to characterize and/or identify a recording location.
We discuss existing methods and their limitations \wrt forensic environment identification in Sec.~\ref{subsec:existing_methods} and Sec.~\ref{subsec:limitations_related_works} and describe our contribution to address these constraints in Sec.~\ref{subsec:contributions}.
 
 \subsection{Existing Methods}
\label{subsec:existing_methods}
Many works address environment identification as a closed-set classification task in a supervised training paradigm~\cite{peters2012name, moore2014room,moore2013roomprints, papayiannis2017discriminative,moore2018room,papayiannis2020end,azimi2021room,baum2022environment}.
Such methods are able to identify a recording location that stems from a known and fixed set of candidate environments. 

Some proposed approaches include traditional classifiers like \glspl{gmm} or \glspl{svm} trained on hand-crafted features that serve as acoustic fingerprints for the respective environments~\cite{peters2012name, moore2014room, moore2018room, papayiannis2017discriminative, baum2022environment,moore2013roomprints}.
Analytic features, however, oftentimes reach their limits in modeling the complexity  of real-world audio conditions, such that \gls{dl} tools have been increasingly explored for various audio forensic tasks~\cite{moussa23_interspeech, xiang2022forensic, baldini2022microphone, chung2020defence}, including environment identification~\cite{azimi2021room, papayiannis2020end}.
Here, Papayiannis~\etal~\cite{papayiannis2020end} show the superiority of deep \gls{crnn} features as opposed to custom, analytic features.
Additionally, Azimi~\etal~\cite{azimi2021room} use the deep features of pre-trained \glspl{nn} for speaker identification as input to an \gls{svm}.

Other works focus on the estimation of specific environment parameters from single channel audio samples~\cite{moore2013room,gamper2018blind, genovese2019blind, gamper2020blind, deng2020online, gotz2022blind,ick2023blind, gotz2023contrastive}.
Moore~\etal~\cite{moore2013room} use analytical models to estimate the geometry of 2-D rectangular rooms while more recent approaches directly regress characteristic environment parameters with \glspl{nn}.
For example, many works target the estimation of RT$_{60}$, \ie, the time until the \gls{air} energy decays by $60$\,db, or the volume of some space~\cite{gamper2018blind, deng2020online, gotz2022blind, ick2023blind, gotz2023contrastive, genovese2019blind}. However, also other parameters were investigated, like C$_{50}$, \ie, the ratio between the first $50$\,ms of a signal and its remaining late energy~\cite{gamper2020blind, gotz2023contrastive}.

\subsection{Limitations of Existing Methods}
\label{subsec:limitations_related_works}
For forensic purposes, existing methods for environment identification are still too constrained in their practical applicability.

First, the proposed supervised closed-set classifiers are natural models for, \eg,  smart home environments, where it is possible to train on the same recording locations that are queried at test time~\cite{azimi2021room, papayiannis2020end}.
However, the forensic task of environment identification is by design a task that has to deal with unknown recording locations at test time. 
After all, the set of potential recording locations depends on the current case at hand.
As a consequence, using supervised classifiers would require to collect extensive training data on a case-by-case basis which is expensive and impractical. 
In addition, the audio signal may stem from none of the potential candidates, such that open-set instead of closed-set classifiers are needed to reject a sample that does not belong to any class.

Second, location identification from estimated environment parameters also has some pitfalls for forensic applications.
It is by design a two-stage process, and as such relatively prone to errors: the parameters themselves are only estimates and subject to possible confusions, overlap, or just estimation errors, and a subsequent environment identification may add further estimation errors. 

Third, the robustness to signal degradations is an underexplored topic in research on environment identification.
Several works experiment on quasi clean audio signals~\cite{peters2012name,	moore2013room, moore2014room,papayiannis2017discriminative, papayiannis2020end,
	baum2022environment}.
Others simulate noise degradations, but use identical distributions in both the training and the test sets~\cite{gamper2018blind, genovese2019blind, gamper2020blind, deng2020online, ick2023blind, moore2018room, azimi2021room}. 

\subsection{Contributions}
\label{subsec:contributions}
With this work, we propose to overcome these limitations with several measures to take a further step towards robust forensic environment identification.

First, we approach the task as a few-shot classification task.
This allows for the identification of recording environments by collecting only few reference audio samples from a set of case-dependent location candidates.
Few-shot learning has already been successfully explored for audio classification to distinguish between signals of different content like, \eg, background music, bird voices or human speech~\cite{wang2021few, heggan2022metaaudio, wolters2020study, zhang2019few}, but not for the identification of audio recording environments.

Second, while we put our primary focus on few-shot learning to directly identify locations, we demonstrate that our framework is capable of environment parameter estimation.
Despite the limits of using environmental parameter estimates for the identification task, they can still unveil important cues about a location.
For example, \edited{estimated parameters can support or question the tool's location prediction and thus assist in the overall process.}
Also, the task may become relevant if the police does not have a hypothesis about candidate locations, but aims to characterize the recording location for further investigations, such as a coarse estimate of the volume of a room.

Third, we specifically set the focus on audio material in forensic investigations which is usually `in-the-wild' data of uncontrolled characteristics. 
We thus evaluate under significant training and test set mismatches, as in practice, the audio signal might be heavily impacted by unknown degradation factors.
We include varying noise distortions which can be expected from arbitrarily
noisy environments or distortions introduced by low-cost acquisition devices.
We also evaluate the practically important case of lossy audio compression and
also re-compression, which both have not been addressed in previous works to
our knowledge.

Additionally, we more closely analyze the impact of recording position mismatches for few-shot classification, since reverberation traces of a location also depend on the position of sound emitters and capturing devices.
So far, some works on closed-set environment classification have evaluated training/test mismatches of microphone positions, however the actual positions are not documented~\cite{baum2022environment, moore2018room}.
Here, we take a further step forward and give first insights into the influence of specific microphone/speaker positions on a custom dataset of synthetic four-sided rooms, where we differentiate between corridor, rectangular and square shaped room categories. 

We denote the proposed framework as ``EnvId'', as an abbreviation for
``environment identification''. We hope that EnvId will set a
new standard for environment identification from data in the wild, and set a baseline for
further research in this direction.

The remainder of the paper is organized as follows. 
Sec.~\ref{sec:technical_background} explains the technical background on reverberant environments and Prototypical Networks~\cite{snell2017prototypical}.
In Sec.~\ref{sec:methods}, we present the proposed EnvId framework and the data generation pipeline for simulating challenging forensic scenarios. 
Sec.~\ref{sec:exp} contains the experimental evaluation of EnvId and Sec.~\ref{sec:conclusion} concludes the work.


\section{\edited{Technical Background}}
\label{sec:technical_background}
\edited{Details on the acoustic modeling of reverberant in-the-wild audio
signals are provided in Sec.~\ref{subsec:rev_envs}, and the concept of
Prototypical Networks is presented in Sec.~\ref{subsec:prototypical_networks}.}

\subsection{\edited{Acoustic Model for Reverberant in-the-Wild Speech}}
\label{subsec:rev_envs}
Consider an audio signal $a(t)$ that propagates from an emitting source into an environment.
During propagation, $a(t)$  can be reflected by various obstacles of different properties that influence the reflection behavior, like the orientation, shape or material of some object.
Such reflections cause the signal to arrive at slightly different times at a recording microphone, similar to closely overlapping echos.
This phenomenon is commonly referred to as reverberation and the reverberation signature of an environment can be measured as an \gls{air} signal $r(t)$.
A reverberant audio signal $s(t)$ is then commonly modeled as convolution 
\begin{equation}
	\label{eq:clean_reverb}
	s(t) = r(t) * a(t)
\end{equation}
of an (ideally anechoic) audio signal $a(t)$ with the captured \gls{air} $r(t)$, which is the de-facto standard model for
reverberation~\cite{peters2012name, moore2014room,papayiannis2017discriminative,moore2018room,papayiannis2020end,azimi2021room,baum2022environment}.
Hence, one can separately record reverberation-free speech signals and
\glspl{air}, and freely combine those signals to simulate reverberant speech from various environments.

Additionally, an environment usually exhibits some kind of background noise, \eg, from nature or urban surroundings.
Also, microphones can introduce noise during the signal capturing process.
Such noise is commonly modeled as an additive  signal $n(t)$ scaled by scalar noise impact $\alpha$~\cite{gamper2018blind, genovese2019blind, gamper2020blind, deng2020online, ick2023blind, moore2018room, azimi2021room}.
The acoustic model can therefore be extended to
\begin{equation}
	\label{eq:reverb_noise}
	\hat{s}(t) = r(t) * a(t) + \alpha\cdot n(t) \enspace.
\end{equation}

\textcolor{black}{Moreover, lossy compression is a recurring challenge for forensic analysts.
Lossy compression might be performed directly by the capturing device, and/or later upon sharing content over the internet, as well as by editing and re-saving operations.
For this reason, we further extend the signal formation to
\begin{equation}
	\label{eq:reverb_noise_compr}
	\minorRev{\tilde{s}(t) = (f^{\mathcal{C}}_1 \circ ... \circ f^{\mathcal{C}}_N) (r(t) * a(t) +  \alpha\cdot n(t))} \enspace,
\end{equation}
where $f^{\mathcal{C}}_n$ with $n \in [1, N]$ describes one out of $N$
lossy compression operations.}

\begin{figure*}[t]
	\centering
	\scalebox{0.74}{\includesvg{images/framework_mr.svg}}
	\caption{\edited{Our end-to-end trainable EnvId framework for joint few-shot environment identification and blind parameter regression from audio recordings. The framework takes audio signals as inputs (a), and consists of a neural feature extractor (b) and projector (c) to process and map the input samples to the learnable, metric embedding space (d). The audio representations in the metric space can both be used for the identification of environments, and the regression of environmental parameters (e).}}\label{fig:envid}
\end{figure*}

\subsection{\edited{Prototypical Networks}}
\label{subsec:prototypical_networks}

Prototypical Networks~\cite{snell2017prototypical} are a representation learning approach that enables few-shot classification, such that a classifier does not have to be retrained for classes that are not part of the training data.
The approach has primarily been used for computer vision tasks, but some studies also demonstrate its effectiveness on audio signals~\cite{wolters2020study, zhang2019few, heggan2022metaaudio, wang2021few}.

The idea of a Prototypical Network~\cite{snell2017prototypical} is to learn a metric embedding space that generalizes to unseen classes. 
In that space, some class is represented by its prototype which is computed from few reference samples.
Some input sample, the query, can then be assigned to a class by measuring the distance between its embedded representation to all classes' prototypes and choosing the closest one.

For the learning approach, the training is split into episodes that follow a $N$-way $K$-shot protocol.
Thus, in each episode, a subset of $N \leq N_t$ classes is randomly sampled from the training set of $N_t$ classes in total.
For each class $c$ in the subset, a support set $\mathcal{S}_c$ of $K$ reference samples $s_r \in \mathcal{S}_c$ is randomly chosen for prototyping, while the remaining samples serve as input queries $s_q \notin \mathcal{S}_c$.
Per class, a prototype $\mathbf{p}_{c}$ is computed as the average vector in the embedding space, formally
\begin{subequations}
	\label{eq:prototype_computation}
	\begin{equation}
		\mathbf{p}_{c} = \frac{1}{K} \sum_{\mathbf{x}_k \in \mathcal{S}_{c}} f(\mathbf{x}_k)\enspace, \mathrm{where}
	\end{equation}
	\begin{equation}
		f : \mathbb{R}^D \rightarrow \mathbb{R}^E
		\label{eq:projector}
	\end{equation}
\end{subequations}
is a learnable function that projects the reference samples' feature vectors $\mathbf{x} \in \mathbb{R}^D$ to the dimension $E$ of the metric space.

Then, the distance, typically the Euclidean distance, between each embedded query sample $\mathbf{q} \in \mathbb{R}^D$ and each prototype $\mathbf{p}_{c}$ is computed as
\begin{equation}
	\operatorname{d}(\mathbf{p}_{c}, f(\mathbf{q})) = \|\mathbf{p}_{c}-f(\mathbf{q})\|\enspace.
	\label{eq:distance}
\end{equation} 

The likelihood that the class label $y$ of a query sample is $c$ is further given by the softmax function over the distances to all classes' prototypes
\begin{equation}
	\label{eq:likelihood}
	p(y = c|\mathbf{q}) = \frac{\operatorname{exp}(		\operatorname{d}(\mathbf{p}_{c}, f(\mathbf{q})))}{\sum_{i = 0}^{N -1} \operatorname{exp}(\operatorname{d}(\mathbf{p}_{i}, f(\mathbf{q})))}\enspace.
\end{equation}
The training loss is the negative log-likelihood
\begin{equation}
	\mathcal{L}_{\mathrm{class}} = -\operatorname{log }p(y = c| \mathbf{q})\enspace,\label{eqn:class_loss}
\end{equation}
which is minimized for the correct class $c$. 

\section{Methods}
\label{sec:methods}

The proposed framework for few-shot environment identification and parameter estimation is presented in Sec.~\ref{subsec:envid_methods}. The associated data generation pipeline is presented in Sec.~\ref{subsec:data_pipeline}.

\subsection{The EnvId Framework}
\label{subsec:envid_methods}

EnvId is designed as an end-to-end trainable framework that handles joint
few-shot location identification and environmental parameter regression.

First, at its core, EnvId \edited{implements the Prototypical Network approach (\cf Sec.~\ref{subsec:prototypical_networks}) to prevent case-specific retraining (\cf Sec.~\ref{subsec:limitations_related_works}).
For our application, the goal is to learn a metric embedding space in which distances enable a distinction of recording locations.
During inference, an audio sample in question (query) and $K$ reference samples from each of $N$ candidate locations can be projected to the metric space.
The query can then be assigned to the closest location prototype via $N$-way $K$-shot classification.}

Second, in case of criminal investigations, some given audio material might not stem from any of the candidate locations.
Thus, EnvId supports open-set few-shot classification, \ie, can reject the query as belonging to an \emph{unknown} class (location), if the distance to all prototype candidates is too large.

Third, the query representations in the metric space are used to support environmental parameter estimation as a side task.
\edited{The blind regression of scalar parameters can both act as an additional cue to the identification result and is still applicable if no candidate set of potential locations is available in a case.}

An overview of the EnvId framework is shown in Fig.~\ref{fig:envid}.
\edited{In Sec.~\ref{subsubsec:envid_inputs} -- Sec.~\ref{subsubsec:envid_outputs}, the framework's individual components and inference process are presented in detail, and Sec.~\ref{subsubsec:envid_training} explains the training procedure.}

\subsubsection{Audio Recording Inputs (Fig.~\ref{fig:envid}\,(a))}\label{subsubsec:envid_inputs}
\minorRev{The input to EnvId consists of the query audio sample $s_q$ to classify and a support set $\mathcal{S}_c$ of $K$ reference audio samples for each of $N$ candidate locations.}
All audio signals are transformed to dense frequency representations, as it is commonly done in audio processing~\cite{salvi2023you, moussa23_interspeech, baldini2022microphone, papayiannis2020end}.
In detail, we compute the Mel spectrogram~\cite{logan2000mel} and the \glspl{mfcc}~\cite{logan2000mel} and 
concatenate the coefficients of both representations into one feature vector.
The Mel spectrograms are computed with torchaudio~\cite{torchaudio} with a window size and FFT size of $1024$ bins, a stride of $512$, and $256$ Mel filter banks.
\edited{This results in a size of $96$ for the time dimension and $256$ for the frequency dimension.}
The \glspl{mfcc} uses 20 coefficients and is computed from the spectrogram. 
\edited{Thus, the final concatenated feature input is of size $96 \times 276$.}

\subsubsection{Feature Extraction (Fig.~\ref{fig:envid}\,(b))}
\label{subsubsec:envid_extractor}
Deep feature vectors of $D$ dimensions are extracted from each input frequency representation.
\minorRev{Here, we recommend a slim \gls{cnn} feature extractor that we refer to as Gamper$^\star$, \ie, a variation of the CNN by Gamper~\emph{et
al.}~\cite{gamper2018blind}: different from the original work, we
use uniform $3\times 3$ filter kernels in all layers instead of large kernel dimensions in the time domain.}
Gamper~\emph{et al.}~\cite{gamper2018blind} aim to capture mostly temporal information from the spectrogram input with a large receptive field in the time domain. 
However, we could not observe a performance gain for weighting temporal over spectral information on our task.
Instead, uniform receptive fields on both dimensions showed to consistently improve the results.

Note that also any other custom end-to-end trainable feature extractor can be
used for this stage, and we demonstrate the performances of various other networks in
the associated evaluation in Sec.~\ref{subsec:extractor_benchmark_exp}.

\subsubsection{Feature Projection (Fig.~\ref{fig:envid}\,(c))}
The feature projector $f$ projects the deep features' dimension $D$ to the metric embedding space's dimension $E$.
For our studies, we use a simple fully connected layer for the mapping.
While the embedding dimension can be varied, we use $E = 256$ throughout all experiment.
In early studies with $2^6 \le E \le 2^9$, we observed that smaller $E$ led to worse embedding space results while
larger $E$ did not show any advantage.

\subsubsection{Embedding and Prototype Computation (Fig.~\ref{fig:envid}\,(d))}\label{subsec:emb_space_opt}
\edited{The projected features are embedded in the metric space.
For each candidate location class $c$, a prototype $\mathbf{p}_c$ is computed from its embedded support set (\cf Eq.~\ref{eq:prototype_computation}).
Then, the Euclidean distance from the embedded query to all prototypes is calculated (\cf Eq.~\ref{eq:distance}).}

\subsubsection{\edited{Location Classification and Parameter Regression (Fig.~\ref{fig:envid}\,(e))}}\label{subsubsec:envid_outputs}
In the final step, the query sample is attributed to some location (Fig.~\ref{fig:envid}\,(e), top) and/or blind environmental parameter estimation is performed from its embedded representation (Fig.~\ref{fig:envid}\,(e), bottom).

During inference, a query can also be rejected if there is no prototype within a reasonably short Euclidean distance.
Otherwise, the likelihood of the query sample belonging to a location is computed from the query-prototype distances in the metric space as defined in Eq.~\ref{eq:likelihood}.
\edited{The final classification is then performed by choosing the closest, \ie, most likely location prototype as
\begin{equation}
	\hat{y} = \underset{c}{\operatorname{argmax}} (p(y=c | \mathbf{q}))\enspace ,
\end{equation}
where $\hat{y}$ is the predicted class label.}

\edited{For the environmental parameter estimation side-task, a trainable function $g : \mathbb{R}^E \rightarrow \mathbb{R}$ regresses a scalar prediction $\hat{p}_e$ from the embedded query $\mathbf{q}$, formally
\begin{equation}
	\hat{p}_e = g(\mathbf{q})\enspace ,
\end{equation}
where we use two linear layers with a hidden dimension of each $256$ for $g(\cdot)$.}

Note that during inference, it is possible to exclusively use EnvId for parameter regression if no set of candidate locations are available for few-shot identification in some forensic case.
\textcolor{black}{In our experiments in Sec.~\ref{subsec:param_est} the regression task is evaluated for the environmental parameters volume and RT$_{60}$, \ie, the time in which the signal energy decays by $60$\,dB.}

\subsubsection{\edited{Training}}\label{subsubsec:envid_training}
The EnvId framework is trained in episodes with a $N$-way $K$-shot protocol (\cf Sec.~\ref{subsec:prototypical_networks}), where the few-shot classification task is optimized using the negative log-likelihood loss $\mathcal{L}_{\mathrm{class}}$ (Eq.~\ref{eq:likelihood}).
If environmental parameter regression is enabled, the metric space (Fig.~\ref{fig:envid} (d)) is optimized to additionally represent scalar environmental characteristics in embedded samples with the regression loss
\begin{equation}
	\mathcal{L}_{\mathrm{reg}} = | p_e - \hat{p}_e |\enspace ,
\end{equation}
which penalizes the absolute deviation of the parameter label $p_e$ from its regressed prediction $\hat{p}_e$.
Naturally, training this side-task requires labeled audio samples with known environmental properties.
For joint training, the total loss
\begin{equation}
	\mathcal{L}_{\mathrm{total}} =  \mathcal{L}_{\mathrm{class}} + \mathcal{L}_{\mathrm{reg}} =
	-\operatorname{log }p(y = c| \mathbf{q}) + 	| p_e - \hat{p}_e |
\end{equation}
is minimized.

\begin{figure*}[t]
	\centering
	\resizebox{\textwidth}{!}{
		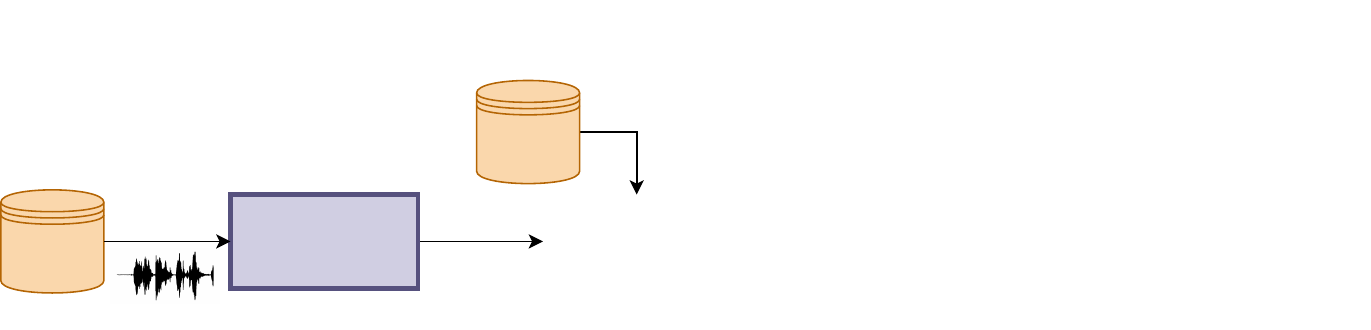}
	\caption{Proposed pipeline for controlled simulation of real world audio recording and post-processing scenarios.
		Configurable sets of input signals, environments and degradations (orange) enable the creation of custom test cases. In $3$ steps (purple), anechoic audio signals $a(t)$ pass various transformations and are output in frequency representation.
		Dashed arrows indicate skip connections to (randomly) enable and disable degradation transformations per sample.}\label{fig:pipeline}
\end{figure*}
\subsection{\edited{Data Generation Framework}}
\label{subsec:data_pipeline}

The quite unconstrained nature of in-the-wild data must be met with extensive training on diverse samples.
Our data samples are hence created with a flexible generation pipeline that is built upon the acoustic model for reverberant speech in Sec.~\ref{subsec:rev_envs}.
The proposed framework is visualized in Fig.~\ref{fig:pipeline}.

In the first step (Fig.~\ref{fig:pipeline}\,(a)), clean reverberant speech samples are simulated from a set of \glspl{air} $r(t)$ and a corpus of anechoic speech samples $s(t)$ according to Eq.~\ref{eq:clean_reverb}.

In the second step, shown in Fig.~\ref{fig:pipeline}\,(b), additive noise degradation is added using Eq.~\ref{eq:reverb_noise}. 
This step is optional and can be skipped depending on the training or evaluation goal for EnvId.
The operation requires a set of noise signals $n(t)$ and an interval of \glspl{snr} that controls the noise strength $\alpha$.
Per reverberant sample, the generator randomly chooses a noise configuration from those specifications.

In the third step, shown in Fig.~\ref{fig:pipeline}\,(c), lossy compression is
introduced as defined in Eq.~\ref{eq:reverb_noise_compr}.
Correspondingly to noise degradation, this step is optional and the specific compression algorithm  $f^{\mathcal{C}}(\cdot)$ with strength $\mathcal{C}$ is randomly sampled from a customisable set for each incoming sample.
Our pipeline natively supports the popular formats MP3, AMR-NB, GSM and
Vorbis, and also the neural EnCodec~\cite{defossez2022highfi} format.
Re-compression is simulated by repeating this random compression process.

The pipeline is configured adaptively to generate data sets for our different training and evaluation scenarios presented in Sec.~\ref{sec:exp}.
The different pipeline settings are described in the respective experiment sections.

\section{Experiments}
\label{sec:exp}
The evaluation section is organized in nine parts.
Initially, Sec.~\ref{subsec:experimental_setup} describes our overall setup for the experiments, Sec.~\ref{subsec:eval_metrics} gives an overview over the reported metrics and Sec.~\ref{subsubsec:data_sources} lists our sample sources for data set creation.
The first of six experiments is described in Sec.~\ref{subsec:extractor_benchmark_exp} which reports EnvId's performance with different feature extractors for the comparison to related work.
Sec.~\ref{subsec:open_set} investigates open-set few-shot classification, and Sec.~\ref{subsec:robustness_deg} explores EnvId's robustness towards unseen signal degradations.
In Sec.~\ref{subsec:k_ablation}, an ablation on the choice of $K$-shot for few-shot inference is provided.
Sec.~\ref{subsec:synth_stuff} presents our study on the influence of specific recording positions in an environment.
Finally, Sec.~\ref{subsec:param_est} explores blind environmental parameter regression at the example of the RT$_{60}$ and the volume.

\subsection{\edited{Experimental Setup}}
\label{subsec:experimental_setup}
We train EnvId with a 10-way 15-shot protocol in episodes (\cf Sec.~\ref{subsec:prototypical_networks}).
Snell et al.~\cite{snell2017prototypical} recommend to set $N$-way and $K$-shot to a value that is expected in practice during inference.
Moreover, they report that training with a larger $N$ can further increase the performance.
Here, we see it as a practically feasible assumption that up to $15$ reference recordings are collected from each of at maximum $10$ candidate locations in a forensic case.
During inference, we set $K=15$ by default, but we also investigate the impact of collecting fewer $K$ reference samples in Sec.~\ref{subsec:k_ablation}, as collecting samples is a relevant cost factor in practice.
Furthermore, results are reported for the most difficult case of $N=10$ candidate classes, or the maximum number of available classes in case of data sets with fewer locations.

Additionally, note that our experiments exclusively evaluate few-shot classification with distinct locations in training and test. 
While the few-shot setting allows for classifying samples from both seen and unseen classes, our experiments only evaluate the identification of environments never seen in training.
This challenging setting results from the nature of forensic investigations that, in practice, deal with highly case-specific candidate locations that are naturally unknown during the preceding training of the identification tool.

Training is conducted using the Adam~\cite{kingma2014adam} optimizer with learning rate $1e^{-4}$ for at maximum $300$ epochs over the training set. 
Early stopping is applied if the validation accuracy for assigning audio samples to recording locations does not increase for $30$ epochs.

Technically, the framework is implemented in PyTorch~\cite{PaszkePyTorch2019} v.1.10.2. Training is conducted on one consumer GPU, at most a RTX3090.

\subsection{\edited{Classification Metrics}}
\label{subsec:eval_metrics}
\edited{We report the accuracy, precision, recall and F1-score to evaluate EnvId's environment identification capabilities.}

\edited{The accuracy is reported over test sets of locations.
It denotes the fraction of query audio input samples that are correctly appointed to their respective recording location.
In some cases, the Top-$n$ accuracies with $n \in [1,3]$ are reported which includes softer accuracy scores where a prediction is counted as correct if the correct location is among the first $n$ predictions.}

\edited{The F1-score, precision and recall are reported for individual locations to analyze EnvId's performance for different types of environments more closely.
As by default, per location, the precision (P), \ie, specificity, is computed by $\frac{T_p}{T_p + F_p}$, where $T_p$ and $F_p$ denote the true positives and false positives, respectively. 
The metric thus reports the correct proportion of all query samples that EnvId appoints to the respective location.
Accordingly, the recall (R), \ie, sensitivity per location is computed by $\frac{T_p}{T_p + F_n}$, where $F_n$ denotes the false negatives. 
This defines EnvId's correctly identified proportion of all query samples from the respective environment.
The F1-score per location summarizes both capabilities in one score, defined as $\frac{2 \cdot P \cdot R}{P + R}$.}

\subsection{\edited{Speech and AIR Resources for Data Set Generation}}
For the simulation of reverberant speech (\cf Sec.~\ref{subsec:data_pipeline}), anechoic single channel audio signals are collected from two freely available databases and  \glspl{air}  are collected from a total of five.
The \gls{air} resources contain real \gls{air} measurements and high-quality \gls{air} simulations that represent a wide range of recording environments.
Tab.~\ref{tab:data_pools} summarizes the source data pools that are further described below.
\textcolor{black}{Note that we additionally use custom synthetic \glspl{air} to perform some evaluations under controlled settings. Those are later described in Sec.~\ref{subsec:synth_stuff} and Sec.~\ref{subsec:param_est}.}
\label{subsubsec:data_sources}

\begin{table}[t]
	\centering
	\caption{Composition of the source data pools of anechoic speech samples and \glspl{air} for data set generation.}\label{tab:data_pools}
	\begin{tabular}{l@{\hskip 1cm}ll}
		\toprule[1pt]
		\multirow{2}{*}{Data Pool} & \multicolumn{1}{c}{Speech $a(t)$} & \multicolumn{1}{c}{AIR $r(t)$} \\ \cmidrule{2-3}
		
		& Source: \# Samples         & Source: \# Samples           \\ \toprule[1pt]
		Training               & ACE~\cite{eaton2016estimation}        :  239                    & MIT~\cite{traer2016statistics}              :  200                   \vspace{0.5em} \\ \vspace{0.5em}
		Validation                 & ACE~\cite{eaton2016estimation}   :  64                     & MIT~\cite{traer2016statistics}               : 20                    \vspace{0.1cm} \\ 
		\multirow{4}{*}{Test}                     & \multirow{4}{*}{TSP~\cite{kabal2002tsp}          : 67}                    & ACE~\cite{eaton2016estimation}               :    7                 \\
		                     &                    & AAIR+REV~\cite{jeub2009binaural,kinoshita2013reverb}       : 13                     \\
		                     &                    & MIT~\cite{traer2016statistics}               :  20                    \\
		                     &                    & OPENAIR~\cite{murphy2010openair}           : 20                   \\ \bottomrule[1pt]
		
	\end{tabular}

\end{table}

\subsubsection{Anechoic Speech Pools.}
Anechoic speech signals $a(t)$ are taken from the ACE~\cite{eaton2016estimation} corpus for training/validation and from the TSP~\cite{kabal2002tsp} corpus for testing.
From the ACE set~\cite{eaton2016estimation}, we use $33$ speech samples from $4$ female speakers and from $7$ male speakers with a duration between $3$\,s and $97$\,s. 
From TSP, we use $87$ utterances from $2$ female speakers and from $2$ male speakers with a duration between $1.70$\,s and $3.25$\,s. 
We train and test EnvId on audio snippets of $3$\,s, so, longer speech samples are split into $3$\,s segments.
TSP also contains shorter samples, which are enlarged by concatenating speech from the same speaker at silent positions.
This results in a total of $239$/$64$ training/validation samples from the
ACE~\cite{eaton2016estimation} corpus, and 
$67$ clean voice snippets from TSP for testing as shown in 
the left column of Tab.~\ref{tab:data_pools}.

\begin{table*}[t]
	\centering
	\newcommand{\hskipper}{0.25cm}
	\caption{Accuracy values (mean $\pm$ standard deviation and maximum) of $5$ training runs per feature extractor on the $4$ test sets of noisy and single-compressed reverberant samples (\cf Sec.~\ref{subsubsec:dataset_benchmark}). The best (bold), second best (underlined) and third best (italic) model results are highlighted. The parameter size includes the feature extractor and EnvId's remaining trainable weights. We provide an own custom extractor (last row) as strong baseline for further research.}\label{tab:cross_data_eval}
		\begin{tabular}
			{@{}l@{\hskip \hskipper}r@{\hskip \hskipper}l@{\hskip \hskipper} l@{\hskip \hskipper}l@{\hskip \hskipper}l@{\hskip \hskipper}l@{\hskip \hskipper}l@{\hskip \hskipper}l@{\hskip \hskipper} l@{\hskip \hskipper}l@{\hskip \hskipper}l@{\hskip \hskipper}l@{\hskip \hskipper}l@{\hskip \hskipper} @{}}
		
			\toprule[1pt]
			
			\multirow{2}{*}{Feature Extractor}&	\multicolumn{1}{l}{\multirow{2}{*}{Params}}&\phantom{l} & \multicolumn{2}{c}{ACE~\cite{eaton2016estimation}}   &&                  \multicolumn{2}{c}{AAIR+REV~\cite{jeub2009binaural,kinoshita2013reverb}}&                      & \multicolumn{2}{c}{MIT~\cite{traer2016statistics}} &                   & \multicolumn{2}{c}{OPENAIR~\cite{murphy2010openair}}                     \\  \cmidrule{4-5} \cmidrule{7-8} \cmidrule{10-11}  \cmidrule{13-14}
			&&& \multicolumn{1}{c}{$\mu\pm\sigma$} & \multicolumn{2}{l}{{$\max$}} & \multicolumn{1}{c}{$\mu\pm\sigma$} & \multicolumn{2}{l}{$\max$} & \multicolumn{1}{c}{$\mu\pm\sigma$} & \multicolumn{2}{l}{$\max$} & \multicolumn{1}{c}{$\mu\pm\sigma$} & \multicolumn{1}{l}{$\max$}  \\ \midrule[1pt]
			Guessing chance             &   && 0.1430              &              && 0.1000              &              && 0.1000            &           && 0.1000          &       \vspace{0.1cm}      \\  
			\textbf{Related Work --- CRNNs:} & \multicolumn{13}{l}{ } \\ 
			Papayiannis~\cite{papayiannis2020end}              &2.66 M&& 0.6768$\scriptstyle\pm0.0314$                & 0.7122               && 0.6866$\scriptstyle\pm0.0166$                & 0.7072                && 0.7134$\scriptstyle\pm0.0068$                & 0.7239              &&0.5652$\scriptstyle\pm0.0274$                & 0.5985                \\
			GamperCRNN~\cite{gamper2020blind}              &1.25 M && 0.7062$\scriptstyle\pm0.0336$                & 0.7676                && 0.7362$\scriptstyle\pm0.0308$                & 0.7738                && 0.7439$\scriptstyle\pm0.0181$                & 0.7761                && 0.5551$\scriptstyle\pm0.0321$                &0.6157                \\ 
			Deng/Götz~\cite{deng2020online,gotz2022blind}             &5.49 M && 0.8068$\scriptstyle\pm0.0271$                & 0.8401               &&0.7614$\scriptstyle\pm0.0564$                & 0.8427                &&0.6755$\scriptstyle\pm0.0647$                &0.7463                &&0.6204$\scriptstyle\pm0.0795$                & 0.7373                 
			\vspace{0.1cm}      \\  
			\textbf{Related Work --- CNNs:} & \multicolumn{13}{l}{ } \\ 
			VGGVox~\cite{nagrani2017voxceleb,azimi2021room}                &11.65 M&& 0.4827$\scriptstyle\pm0.0079$                 & 0.4947                && 0.5362$\scriptstyle\pm0.0184$                 & 0.5637                 && 0.5497$\scriptstyle\pm0.0065$                 & 0.5567                 && 0.4727$\scriptstyle\pm0.0201$                 & 0.5037                 \\
			Genovese/Ick~\cite{genovese2019blind,ick2023blind}        &0.19 M&&0.8196$\scriptstyle\pm0.1028$                 & 0.8721                && 0.7715$\scriptstyle\pm0.0841$                 & 0.8220               &&0.7340$\scriptstyle\pm0.1149$                 & 0.8022              &&0.6937$\scriptstyle\pm0.1064$                 & 0.7448                \\
			ThinResnet~\cite{xie2019utterance,azimi2021room}                &11.72 M&&0.8499$\scriptstyle\pm0.0178$                 & 0.8806                 && 0.8583$\scriptstyle\pm0.0190$                 & 0.8783                 && \emph{0.8257}$\scriptstyle\pm0.0054$                 & 0.8328                 &&0.6907$\scriptstyle\pm0.0329$                 & 0.7388                 \\
			Götz~\cite{gotz2023contrastive}               &3.86 M&& \emph{0.9416}$\scriptstyle\pm0.0218$                 & 0.9616                && \emph{0.9150}$\scriptstyle\pm0.0355$                 & 0.9518              &&0.8236$\scriptstyle\pm0.0454$                 & 0.8679                &&\underline{0.8187}$\scriptstyle\pm0.0501$                 & 0.8754                 \\
			GamperCNN~\cite{gamper2018blind}                &3.43 M&&\underline{0.9467}$\scriptstyle\pm0.0189$                 & 0.9638                 &&\underline{0.9254}$\scriptstyle\pm0.0291$                 & 0.9506                 && \underline{0.8345}$\scriptstyle\pm0.0419$                 & 0.8687                 && \textbf{0.8240}$\scriptstyle\pm0.0486$                 & 0.8709                    \vspace{0.1cm}      \\  
			\textbf{Standard Vision CNNs:} & \multicolumn{13}{l}{ } \\ 

			RegNetY-400mf~\cite{radosavovic2020designing}                &28.23 M&& 0.7211$\scriptstyle\pm0.0441$                 & 0.7868                 && 0.7724$\scriptstyle\pm0.0307$                 & 0.8129                 && 0.7334$\scriptstyle\pm0.0387$                 & 0.7851                &&0.6163$\scriptstyle\pm0.0400$                 & 0.6649                \\				
			EffNet-B0~\cite{tan2019efficientnet}                &12.85 M&&0.7318$\scriptstyle\pm0.0303$                 & 0.7868                && 0.8101$\scriptstyle\pm0.0157$                 & 0.8301                 && 0.7939$\scriptstyle\pm0.0116$                 & 0.8067                && 0.6300$\scriptstyle\pm0.0191$                 & 0.6642                \\
			
			EffNet-B2~\cite{tan2019efficientnet}                &17.43 M&&0.7548$\scriptstyle\pm0.0199$                 & 0.7932                 && 0.8057$\scriptstyle\pm0.0278$                 & 0.8542                 && 0.8078$\scriptstyle\pm0.0121$                 & 0.8291                &&0.6639$\scriptstyle\pm0.0209$                 & 0.6925                 \\
			RegNetY-800mf~\cite{radosavovic2020designing}                &43.75 M&& 0.7646$\scriptstyle\pm0.0187$                 & 0.7846               && 0.7929$\scriptstyle\pm0.0114$                 & 0.8140                 &&0.7640$\scriptstyle\pm0.0213$                 & 0.7888                 && 0.6436$\scriptstyle\pm0.0141$                 & 0.6604                 \\
			ResNet-18~\cite{he2016deep}                &11.30 M&& 0.8171$\scriptstyle\pm0.0345$                 & 0.8593                 && 0.8200$\scriptstyle\pm0.0276$                 & 0.8588                 && 0.8124$\scriptstyle\pm0.0111$                 & 0.8269                && 0.6778$\scriptstyle\pm0.0348$                 & 0.7336                 \\
			ResNet-50~\cite{he2016deep}                &24.03 M&&0.8175$\scriptstyle\pm0.0113$                 & 0.8316                 && 0.8361$\scriptstyle\pm0.0110$                 & 0.8485                && 0.8045$\scriptstyle\pm0.0106$                 & 0.8216                 && 0.6840$\scriptstyle\pm0.0076$                 & 0.6925                \\
			ConvNeXt-Tiny~\cite{liu2022convnet}               &28.01 M&& 0.8618$\scriptstyle\pm0.0112$                 & 0.8806                && 0.7986$\scriptstyle\pm0.0137$                 & 0.8129                 && 0.7658$\scriptstyle\pm0.0190$                 & 0.8000                 && 0.7022$\scriptstyle\pm0.0118$                 & 0.8000  \vspace{0.1cm}\\
			\textbf{Standard Transformers:} & \multicolumn{13}{l}{} \\ 
			Transf. Enc.~\cite{vaswani2017attention}                &19.19 M&&0.4034$\scriptstyle\pm0.0394$                 & 0.4670                && 0.4693$\scriptstyle\pm0.0388$                 & 0.5385                 &&0.4663$\scriptstyle\pm0.0141$                 & 0.4903                 && 0.4227$\scriptstyle\pm0.0227$                 & 0.4545                 \\
			ViT-Small~\cite{dosovitskiyimage}                &86.08 M&& 0.4968$\scriptstyle\pm0.0350$                 & 	0.5437                 &&0.5146$\scriptstyle\pm0.0234$                 & 0.5408                 && 0.4922$\scriptstyle\pm0.0113$                 & 0.5045                 && 0.4813$\scriptstyle\pm0.0128$                 & 0.5015               
			\vspace{0.1cm}\\
			\textbf{Proposed Backbone:} & \multicolumn{13}{l}{} \\ 
			Gamper$^\star$                &3.86 M&& \textbf{0.9557}$\scriptstyle\pm0.0076$                 & 	0.9701                 &&\textbf{0.9375}$\scriptstyle\pm0.0047$                 &0.9449                 && \textbf{0.8712}$\scriptstyle\pm0.0164$ & 0.8903 &&\emph{0.8131}$\scriptstyle\pm0.0188$    & 0.8358		               
			\\		\bottomrule[1pt]   
		\end{tabular}

\end{table*}

\subsubsection{\gls{air} Pools.}
The training/validation pools of \gls{air} signals $r(t)$ consist of real $200$/$20$ \glspl{air} from the MIT Acoustical Reverberation Scene Statistics Survey~\cite{traer2016statistics} (MIT) data set.
The MIT set provides diverse space categories.
Small categories are, \eg, \emph{cars} and \emph{bathroom}, mid-size categories are, \eg, \emph{bar}, \emph{train}, and \emph{hallway}, and large categories are, \eg, \emph{theatre}, \emph{atrium}, or \emph{open air}.
For testing we use four separate \gls{air} pools from the remaining MIT \glspl{air} and from smaller databases that are described below.

The ACE test pool consists of the ACE \gls{air} set of $7$ rooms that is oftentimes used in related work~\cite{ick2023blind, gotz2022blind, azimi2021room, deng2020online, gamper2020blind, papayiannis2020end, genovese2019blind, gamper2018blind, moore2018room, baum2022environment}.
The rooms are enclosed mid-size spaces of the category \emph{lecture room},
\emph{meeting room}, \emph{office room} and \emph{lobby}, with volumes between
$47.3$\,m$^3$ and $370$\,m$^3$. 

The AAIR+REV test pool is created by combining the $7$ real \gls{air} measurements from the Aachen Impulse Response Database (AAIR)~\cite{jeub2009binaural} with $6$ high-quality simulations from the 2014 REVERB challenge (REV)~\cite{kinoshita2013reverb}.
AAIR provides measurements from a low-reverberant studio booth, a stairway, an office, a meeting room, two lecture rooms, and a church. It features diverse surface materials and furniture. 
The room volumes range between $11.9$\,m$^3$ and $370.8$\,m$^3$, except for the church where only the floor area is reported with $570$\,m$^2$.
To enlarge the data set, we sample $6$ high-quality simulations from the REV set following Kinoshita~\etal~\cite{kinoshita2013reverb}, where $2$ rooms are from each category \emph{small}, \emph{medium} and \emph{large}. 

The MIT and OPENAIR test pools contain $20$ \gls{air} measurements each.
The MIT~\cite{traer2016statistics} database provides \glspl{air} from variable space categories while the OPENAIR~\cite{murphy2010openair} data mainly covers large spaces like churches, halls, auditoriums and open spaces, where the volumes (excluding open air) range between $35.2$\,m$^3$ and $140000$\,m$^3$.

For more details, we list the specific selected environments later in Tab. III for the OPENAIR pool. Analogously, the selected environments from the ACE, AAIR+REV and MIT test pools are listed in Tab.~A.1, A.2 and A.3 in the supplemental material.\footnote{\url{https://faui1-files.cs.fau.de/public/mmsec/moussa/2025TIFS/supplemental_material.pdf}}

\subsection{Feature Extractor Benchmark for Few-Shot Identification}
\label{subsec:extractor_benchmark_exp}

We train \gls{nn} models from related work and popular standard architectures as feature extractors of the EnvId framework (Fig.~\ref{fig:envid} (b)) to be able to compare them to the proposed EnvId configuration (\cf Sec.~\ref{subsubsec:envid_extractor}) on our task of forensic few-shot location identification.

\edited{The experiment is presented in four parts.
First, we define the experimental setup, where Sec.~\ref{subsubsec:dataset_benchmark} presents the generated training and test data sets and Sec.~\ref{subsubsec:feature_extractors} summarizes the feature extractors for evaluation.
Then, the experimental results are discussed in Sec.~\ref{subsubsec:feature_ex}, and Sec.~\ref{subsubsec:gen_ability} provides an analysis of the generalization ability towards identifying out-of-distribution environments.}

\subsubsection{\edited{Training and Test Data Sets}}
\label{subsubsec:dataset_benchmark}
The data generation pipeline from Sec.~\ref{subsec:data_pipeline} is used to generate a training set, a validation set and four test sets from the respective speech and \gls{air} pools in Tab.~\ref{tab:data_pools}.
Note that for future reference, we name the test sets after their underlying environment pools, \ie, as ACE, AAIR+REV, MIT and an OPENAIR test set.
To create a set, each sample from the speech pool is convolved with each sample from the \gls{air} pool, which leads to a uniform distribution of recording examples for each available environment (Fig.~\ref{fig:pipeline}\,(a)).
This experiment features the same common signal degradations in the training and test sets, while robustness towards unseen degradations is evaluated separately in Sec.~\ref{subsec:robustness_deg}.
In detail, for each sample in a set, general background noise is simulated with additive white noise of random strength within a broad \gls{snr} range of $\alpha \in [-10, 50] \cup [\infty]$\,db, where $\alpha = \infty$ corresponds to no noise being added (Fig.~\ref{fig:pipeline}\,(b)). 
Single-compression (Fig.~\ref{fig:pipeline}\,(c)), is then randomly applied with one of the popular formats MP3, AMR-NB, GSM with arbitrary and supported compression strength, or skipped. 
In detail, the set of available bitrate configurations to the pipeline are $\mathcal{C}_{MP3} = \{8, 16, 24,  32, 40, 48, 56, 64, 80, 96, 112, 128\}$, $\mathcal{C}_{AMR-NB} = \{4.75, 5.15, 5.9, 6.7, 7.4, 7.95, 10.2, 12.2\}$ and $\mathcal{C}_{GSM} = \{13\}$, since GSM only operates with constant bitrate. 

\subsubsection{\edited{Baseline Feature Extractors}}
\label{subsubsec:feature_extractors}
The comparison includes in total $17$ feature extractors from the literature.
This covers $8$ neural feature extractors from related work (\cf Sec.~\ref{subsec:existing_methods}) that we implement from $11$ corresponding published works.
Three extractors are based on CRNNs, which we denote as Papayiannis~\cite{papayiannis2020end}, GamperCRNN~\cite{gamper2018blind}, and	Deng/G{\"o}tz~\cite{deng2020online,gotz2022blind}. 
Five extractors are based on CNNs, which we denote as VGGVox~\cite{nagrani2017voxceleb,azimi2021room}, Genovese/Ick~\cite{genovese2019blind,ick2023blind}, ThinResnet~\cite{xie2019utterance,azimi2021room}, G{\"o}tz~\cite{gotz2023contrastive}, and GamperCNN~\cite{gamper2020blind}.
Additionally, $9$ standard \glspl{cnn} from the RegNet~\cite{radosavovic2020designing}, EffNet~\cite{tan2019efficientnet}, ResNet~\cite{he2016deep} and ConvNext~\cite{liu2022convnet} families and two standard Transformer models~\cite{vaswani2017attention, dosovitskiyimage} are considered, because standard architectures oftentimes show good learning capabilities in various forensic and steganalytic tasks.

Each of the evaluated models  is integrated into EnvId in two steps.
The final classification/regression layers (if present) are removed, and the remaining layers (Fig.~\ref{fig:envid}\,(b)) are linked to EnvId's projector layer (Fig.~\ref{fig:envid}\,(c)). 
The projector reduces the dimensionality of the model to $E = 256$, the size of EnvId's metric embedding space (\cf Sec.~\ref{subsec:envid_methods}).
Missing parameter configurations in related methods are supplemented with the respective standard values from PyTorch~\cite{PaszkePyTorch2019}.

\subsubsection{Comparison of Feature Extractors}\label{subsubsec:feature_ex}
Tab.~\ref{tab:cross_data_eval} shows the evaluation results on all four test sets.
For a representative comparison, we conduct $5$ EnvId training runs with random weight initialization for each feature extractor and report the mean accuracy and standard deviation, as well as the best model accuracy per test set. 
The first row lists the random guessing chance per test set for reference.
Since we report $10$-way few-shot results (\cf Sec.~\ref{subsec:experimental_setup}), this value corresponds to $0.1000$, except for the small ACE set of $7$ locations which enforces a $7$-way setting with higher guessing chance of $0.1430$.
The second and following rows list the results per feature extractor and test set grouped by architecture type and origin, \ie, related work vs. standard architectures.
The rows in a group are sorted by the performance in ascending order. 
Our proposed backbone, \ie, the variation of GamperCNN that we denote as Gamper$^\star$ (\cf Sec.~\ref{subsec:envid_methods}) is shown in the last line.

Several insights can be drawn from these results.
\edited{Overall, the proposed Gamper$^\star$ performs best with an average accuracy score of $0.8944$ over all sets and quite low standard deviations between $0.0047$ and $0.0188$ over the $5$ training runs.
The feature extractors GamperCNN and Götz perform second and third best with average accuracy values of $0.8826$ and $0.8747$ over all sets.
Both yield less stable results over training runs than Gamper$^\star$ with standard deviations between $0.0189$ and $0.0486$ for GamperCNN and between $0.0218$ and $0.0501$ for Götz.
}
Notably, these three \gls{cnn} feature extractors are relatively small, with less than $4$ million parameters each.
Nevertheless, they outperform the much more complex \gls{crnn} extractors and standard models that consist of a much larger number of parameters.
It is particularly notable that the Transformer models achieve only a quite low performance.

We hypothesize that the benefits from large-scale vision nets and Transformers might potentially become accessible with even larger training data sets.
However, due to the very promising results with efficient and slim \glspl{cnn} models with short training times, we did not further investigate this direction.

Detailed performance results for each location class are listed for the ACE, AAIR+REV and MIT set in Tab.~A.1, A.2 and A.3 of the supplemental material, for the proposed Gamper$^\star$ feature extractor and the best baseline models Götz and GamperCNN.\textsuperscript{2}
The reports include the F1-score, precision, and recall for all individual recording locations.
The corresponding scores on the most challenging OPENAIR set are separately discussed below.

\subsubsection{Generalization to Out-of-Distribution Locations}
\label{subsubsec:gen_ability}
The results in Tab.~\ref{tab:cross_data_eval} show that OPENAIR is the most difficult test set.
OPENAIR consists mainly of large, strongly reverberant locations like churches, cathedrals and halls.
Such types of locations are only sparsely covered in the training set.
In this section, we further characterize that sparsity and the associated challenge to generalize to out-of-distribution environments during few-shot classification.

One quantitative indication for the difficulty to generalize to the OPENAIR samples is provided by the Pearson correlation coefficient between the \glspl{air} in the training and test sets.
In detail, the Pearson correlation coefficients for the ACE, AAIR+REV, and MIT test sets
are $0.5555$, $0.3162$ and $0.5844$, whereas it is significantly lower for the OPENAIR test set with a value of $0.2375$.

\begin{table*}[t]
	\centering
		\caption{\edited{Mean F1-score, precision and recall  per environment over $5$ training runs of our EnvId framework on the OPENAIR~\cite{murphy2010openair} test set of noisy and single-compressed reverberant samples (\cf Sec.~\ref{subsubsec:dataset_benchmark}). The results are reported for the Gamper$^\star$, GamperCNN~\cite{gamper2018blind} and Götz~\cite{gotz2023contrastive} feature extractor. }}\label{tab:openair_class_perf}

		\newcommand{\hskipper}{0.14cm}
		\begin{tabular}{@{}l@{\hskip\hskipper}r@{\hskip\hskipper}l@{\hskip\hskipper}l@{\hskip\hskipper}l@{\hskip\hskipper}l@{\hskip\hskipper}l@{\hskip\hskipper}l@{\hskip\hskipper}l@{\hskip\hskipper}l@{\hskip\hskipper}l@{\hskip\hskipper}l@{\hskip\hskipper}l@{\hskip\hskipper}l@{\hskip\hskipper}@{}}
		\toprule[1pt]
		\multirow{2}{*}{OPENAIR Class Label}                       & 	\multirow{2}{*}{Volume}   & 	\multirow{2}{*}{Space Category}&\multicolumn{3}{c}{Gamper$^\star$}&&\multicolumn{3}{c}{GamperCNN}&&\multicolumn{3}{c}{Götz} \\
		\cmidrule{4-6}\cmidrule{8-10}\cmidrule{12-14}
     	&&& F1 & P &  R  && F1 & P & R  && F1 & P & R\\
		\midrule[1pt]
		Innocent Railway Tunnel Entrance             & 13000 m$^3$  & -                           & 0.9664 & 0.9877 & 0.9463 && 0.9548 & 0.9639 & 0.9463 && 0.9565 & 0.9591 & 0.9552 \\
		Live Room                                    & 35.2 m$^3$   & Recording Studio            & 0.9604 & 0.9783 & 0.9433 && 0.9565 & 0.9589 & 0.9552 && 0.9704 & 0.9753 & 0.9672 \\
		
		Falkland Palace Bottle Dungeon               & -         & Chamber                     & 0.9548 & 0.9968 & 0.9164 && 0.9588 & 0.9781 & 0.9403 && 0.9307 & 0.9598 & 0.9045 \\		
	
		Koli National Park Summer                    & -         & Open Air                    &0.9424 & 0.9575 & 0.9284 && 0.9263 & 0.9891 & 0.8716 && 0.9182 & 0.9273 & 0.9104 \\
		
		Stairwell                                    & -         & Hall                        & 0.8998 & 0.8874 & 0.9134 && 0.8633 & 0.8532 & 0.8746 && 0.8892 & 0.8932 & 0.8866 \\
		
		Arthur Sykes Rymer Audit. Univ. York & 1560 m$^3$   & Auditorium                  & 0.8911 & 0.8662 & 0.9224 && 0.8543 & 0.8160 & 0.9015 && 0.8466 & 0.8266 & 0.8687 \\
		
		Spokane Woman's Club                          & 1600 m$^3$   & Auditorium, Ballroom, Hall  & 0.8635 & 0.8488 & 0.8806 && 0.9084 & 0.8891 & 0.9313 && 0.8954 & 0.8702 & 0.9254 \\
		
		Falkland Tennis Court                        & 2300 m$^3$   & Open Air, (Sports) Hall & 0.8610 & 0.8870 & 0.8388 && 0.8069 & 0.8106 & 0.8060 && 0.8335 & 0.8407 & 0.8269 \\

		Dixon Studio Theatre Univ. York          & 908.23 m$^3$ & Theater                     &0.8525 & 0.8049 & 0.9075 && 0.8865 & 0.9030 & 0.8716 && 0.9071 & 0.8992 & 0.9164 \\
		
		Council Chamber                               & 1140 m$^3$   & Chamber                     &0.8524 & 0.8189 & 0.8896 && 0.8196 & 0.7936 & 0.8478 && 0.8374 & 0.8227 & 0.8537 \\
		
		Lady Chapel St Alban's Cathedral              & -         & Cathedral                   & 0.8153 & 0.7733 & 0.8627 && 0.8345 & 0.7910 & 0.8836 && 0.8234 & 0.8106 & 0.8388 \\
		
		Jack Lyons Concert Hall Univ. York      & -         & Concert Hall                & 0.8135 & 0.8052 & 0.8239 && 0.7977 & 0.7846 & 0.8149 && 0.7818 & 0.7622 & 0.8030 \\
		
		Central Hall Univ. York                 & 8000 m$^3$   & Auditorium, Hall             & 0.8047 & 0.7202 & 0.9134 && 0.8514 & 0.8341 & 0.8716 && 0.8577 & 0.8242 & 0.8955 \\

		Alcuin College Univ. York               & 21000 m$^3$  & Open Air                    & 0.7947 & 0.7252 & 0.8806 && 0.8530 & 0.8593 & 0.8478 && 0.8425 & 0.8459 & 0.8418 \\
		
		Baptist Nashville Balcony                    & -         & Church                      & 0.7780 & 0.7904 & 0.7672 && 0.7991 & 0.8287 & 0.7731 && 0.7612 & 0.7879 & 0.7403 \\

		Heslington Church                            & 2000 m$^3$   & Church                      & 0.7689 & 0.8170 & 0.7284 && 0.7874 & 0.8261 & 0.7552 && 0.7919 & 0.8211 & 0.7701 	\vspace{0.2cm} \\
		
		Nuclear Reactor Hall                         & 3500 m$^3$   & Hall                        & 0.6832 & 0.6431 & 0.7313 && 0.6356 & 0.5547 & 0.7493 && 0.6258 & 0.5591 & 0.7134		
		\\
		
		Terry's Warehouse                             & 4500 m$^3$   & Hall                        & 0.5982 & 0.6904 & 0.5343 && 0.6449 & 0.6928 & 0.6060 && 0.6481 & 0.6693 & 0.6299 \\
		
		York Minster                                 & 140000 m$^3$ & Cathedral                   & 0.5592 & 0.7172 & 0.4597 && 0.7374 & 0.8515 & 0.6507 && 0.7435 & 0.8391 & 0.6687 \\

		Sportscentre                                 & 9000 m$^3$   & (Sports) Hall            & 0.5263 & 0.5939 & 0.4746 && 0.6180 & 0.6606 & 0.5821 && 0.5016 & 0.5597 & 0.4567 \\
		
		\bottomrule[1pt]  
	\end{tabular}

\end{table*}

\edited{Especially at the level of individual \glspl{air}, it is possible to observe the
impact of differences in the training and test distributions. 
To this end, we limit the examination to the three best performing backbones Gamper$^\star$, GamperCNN and Götz, and examine the F1-score for each location in the OPENAIR test set.
The averaged performance scores over $5$ training runs per model are listed in Tab.~\ref{tab:openair_class_perf} in descending order w.r.t. the F1-score of the proposed Gamper$^\star$ backbone.
All three models perform worst on the same $4$ cathedral and hall environments (bottom part of Tab.~\ref{tab:openair_class_perf}). 
For these $4$ worst environments, the F1-score is only $0.5917$,
$0.6590$, and $0.6298$ for Gamper$^\star$, GamperCNN and Götz.
Without these $4$ locations, the F1-scores for the remaining locations are much higher at $0.8637$, $0.8661$, and $0.8652$, respectively.
}

\edited{The proposed Gamper$^\star$ performs slightly worse than GamperCNN and
G{\"o}tz on OPENAIR locations, which shows the contradicting goals of specialization and
generalization. 
However, the results are overall quite robust, given that the
tested \glspl{air} are considerably different from the training data.
}

\begin{figure*}[t]
	\centering
	\newcommand{\rocscale}{0.258}
	\newcommand{\rocskip}{\hspace{-0.3cm}}
	\hspace{-0.3cm}
	\subfloat[ACE~\cite{eaton2016estimation} \label{fig:roc_ace}]{	\resizebox{\rocscale\textwidth}{!}{\includesvg{plots/rocs_ace.svg}}}
	\rocskip
	\subfloat[AAIR+REV~\cite{jeub2009binaural,kinoshita2013reverb} \label{fig:roc_aair_rev}]{	\resizebox{\rocscale\textwidth}{!}{\includesvg{plots/rocs_aachen.svg}}}\rocskip
	\subfloat[MIT~\cite{traer2016statistics} \label{fig:roc_mit}]{	\resizebox{\rocscale\textwidth}{!}{\includesvg{plots/rocs_mit.svg}}}\rocskip
	\subfloat[OPENAIR~\cite{murphy2010openair} \label{fig:roc_openair}]{	\resizebox{\rocscale\textwidth}{!}{\includesvg{plots/rocs_openair.svg}}}\rocskip
	\caption{\edited{\Gls{roc} curves on our $4$ test sets of noisy and single compressed reverberant speech from Sec.~\ref{subsubsec:dataset_benchmark} for rejecting samples that do not match any reference recording location.The results are reported for the Gamper$^\star$, GamperCNN~\cite{gamper2018blind} and Götz~\cite{gotz2023contrastive} feature extractor.}}\label{fig:roc}
\end{figure*}

\subsection{\edited{Open-Set Matching: Rejection of Unknown Environments}}
\label{subsec:open_set}
In several important practical use cases, it is helpful if a system can indicate that an input does not match any of the reference locations. 
To our knowledge, this task is not addressed in related works.
EnvId can be straightforwardly used for open-set few-shot classification by rejecting an input query if its distance to all available prototypes exceeds a threshold.
The specific choice of rejection threshold depends on practical requirements, \ie, whether a higher recall (\gls{tpr}) or lower \gls{fpr} is targeted.

\edited{For the evaluation, we only allow a test query's recording location as reference candidate in $50\%$ of the cases.
Consequently, EnvId needs to reject half of the query samples as unknown.
To show this general rejection ability, we calculate the average \gls{roc} over the $5$ trained EnvId checkpoints with feature extractors Gamper$^\star$, GamperCNN and Götz from Tab.~\ref{tab:cross_data_eval} on our $4$ test sets of noisy and single compressed speech samples (\cf Sec.~\ref{subsubsec:dataset_benchmark}).
}

\edited{
The \gls{roc} results together with the \gls{auc} score are plotted in Fig.~\ref{fig:roc}.
The \gls{tpr} denotes the amount of samples correctly rejected as unknown and the \gls{fpr} denotes the amount of samples wrongfully assigned to one candidate location. 
The proposed Gamper$^\star$ feature extractor performs best across all test sets and has the lowest standard deviations over all runs.
Götz and GamperCNN perform second and third best with \gls{auc} values that are on average $7.41$ \gls{pp} and $8.67$ \gls{pp}  behind Gamper$^\star$'s performance.
In line with the results in Tab.~\ref{tab:cross_data_eval}, the best performance  for all models is achieved on the ACE test set (Fig.~\ref{fig:roc_ace}), followed by the AAIR+REV set (Fig.~\ref{fig:roc_aair_rev}).
More in detail, the \gls{auc} score for Gamper$^\star$ is $0.92$ on ACE locations and $0.89$ on AAIR+REV locations.
The difference to the remaining feature extractors is largest for those two sets, where Götz yields scores of $0.83$ and $0.81$ and GamperCNN achieves \gls{auc} values of $0.80$ and $0.79$.
Recognizing unknown locations from queries of the MIT and OPENAIR set is significantly harder (\cf Fig.~\ref{fig:roc_mit} vs. Fig.~\ref{fig:roc_openair}).
Here, the \gls{auc} scores for Gamper$^\star$ are $0.81$ and $0.84$. 
So, rejecting query samples from the OPENAIR set is slightly easier for the classifier, even if the identification scores on these out-of-distribution environments is lower (\cf Sec.~\ref{subsubsec:gen_ability}).
The same is true for Götz and  GamperCNN with in total lower \gls{auc} scores, \ie, \gls{auc} $=0.75$ and \gls{auc} $=0.78$ for Götz and \gls{auc} $= 0.72$ and \gls{auc} $=0.79$ for GamperCNN.
}

\edited{
To sum up, these findings show that EnvId is able to perform open-set few-shot identification of recording environments, but nevertheless there is still room for improvement.
Therefore, we regard the matter as a possibility for future research.
}

\begin{figure*}[t]
	\newcommand{\imgskip}{\hspace{-0.05cm}}
	\hspace{0.1cm}
	\begin{subfigure}{1\textwidth}
		\resizebox{0.33\textwidth}{!}{\includesvg{plots/multi_compr_single_bar.svg}}
		\resizebox{0.33\textwidth}{!}{\includesvg{plots/multi_compr_double_bar.svg}}\imgskip
		\resizebox{0.33\textwidth}{!}{\includesvg{plots/multi_compr_triple_bar.svg}}\imgskip
		
		\vspace{-1.3\baselineskip}
		
		\caption{Multi Compression}\label{subfig:multi_c}
	\end{subfigure}
	
	\hspace{0.1cm}
	\begin{subfigure}{1\textwidth}
		\resizebox{0.33\textwidth}{!}{\includesvg{plots/vorbis_br_78_bar.svg}}
		\resizebox{0.33\textwidth}{!}{\includesvg{plots/vorbis_br_64_bar.svg}}\imgskip
		\resizebox{0.33\textwidth}{!}{\includesvg{plots/vorbis_br_16_bar.svg}}\imgskip
		\vspace{-1.3\baselineskip}
		\caption{Vorbis Compression}\label{subfig:vorbis}
	\end{subfigure}
	
	\hspace{0.1cm}
	\begin{subfigure}{1\textwidth}
		\resizebox{0.33\textwidth}{!}{\includesvg{plots/encodec_br_24_bar.svg}}
		\resizebox{0.33\textwidth}{!}{\includesvg{plots/encodec_br_6_bar.svg}}\imgskip
		\resizebox{0.33\textwidth}{!}{\includesvg{plots/encodec_br_1_5_bar.svg}}\imgskip
		\vspace{-1.3\baselineskip}
		\caption{EnCodec Compression}\label{subfig:encodec}
	\end{subfigure}
	
	\hspace{0.1cm}
	\begin{subfigure}{1\textwidth}
		\resizebox{0.33\textwidth}{!}{\includesvg{plots/noise_snr_50_bar.svg}}
		\resizebox{0.33\textwidth}{!}{\includesvg{plots/noise_snr_25_bar.svg}}\imgskip
		\resizebox{0.33\textwidth}{!}{\includesvg{plots/noise_snr_0_bar.svg}}\imgskip
		\vspace{-1.1\baselineskip}
		\caption{Unseen Real Noise}\label{subfig:noise}
	\end{subfigure}
	
	\caption{\edited{Averaged Top-\{1,2,3\} accuracy of 5 training runs for few-shot environment identification under degradation factors unseen during training. We provide benchmarks for the Gamper$^\star$, GamperCNN~\cite{gamper2018blind} and Götz~\cite{gotz2023contrastive} backbones on the MIT~\cite{traer2016statistics} test set for multi compression runs (Fig.~\ref{subfig:multi_c}), high, mid and low quality settings of the unseen Vorbis (Fig.~\ref{subfig:vorbis}) and neural EnCodec~\cite{defossez2022highfi} compression codecs (Fig.~\ref{subfig:encodec}), and unseen real environmental background noise (Fig.~\ref{subfig:noise})}. 
	\minorRev{Note that the y-axis is scaled differently for the individual degradation types to better highlight variations within each experiment.}
	}
	\label{fig:multi_compression}
\end{figure*}

\subsection{\edited{Robustness towards Unseen Signal Degradations}} \label{subsec:robustness_deg}
\edited{This section provides experiments to characterize EnvId's robustness towards signal degradations unseen during training.
After all, unseen post-processing has to be expected for unconstrained in-the-wild audio data in practice.
We here include unseen multiple compression, compression codecs and background noise.} 

\edited{Equally to the previous evaluations, the experiments are performed with the trained EnvId checkpoints wrapping the three best performing feature extractors Gamper$^\star$, GamperCNN and Götz from Sec.~\ref{subsec:extractor_benchmark_exp}.
Recall that those models were trained on samples degraded by additive white noise of random strength and one compression run with random format and bitrate setting (\cf Sec.~\ref{subsubsec:dataset_benchmark}).
For the evaluation, we use variants of the MIT test set from Sec.~\ref{subsubsec:dataset_benchmark} with altered noise and compression degradation settings as described in the respective sections below.
The MIT set is one of our larger sets and  contains a diverse distribution of recording locations from small to large environments.
}

\subsubsection{Multiple Compression Runs}
When audio messages are shared over the internet, the signal might be recompressed multiple times.
\edited{To analyze such situations, the data pipeline is configured to construct two MIT test set versions, one with double compressed samples and one with triple compressed samples. 
The sets do not contain noise degradation and the compression configurations are sampled from the training distribution to isolate the influence of multiple runs.
In detail, for each of $n \in \{2,3\}$ compression runs per sample, the format and strength is randomly chosen from MP3, AMR-NB and GSM with valid bitrate settings (\cf Sec.~\ref{subsubsec:dataset_benchmark}).}

\edited{Fig.~\ref{subfig:multi_c} shows the average Top-\{1,2,3\} accuracy over $5$ training runs per model for single, double and triple compression.
The proposed Gamper$^\star$ feature extractor performs best in all cases, followed by GamperCNN and Götz.
As expected, the accuracy degrades with increasing number of compressions for all backbones.
However, the total performance remains high even for Top-1 predictions. 
For Gamper$^\star$, the accuracy is $0.9176$ for double compression and $0.8937$ for triple compression, which is only moderately lower than  $0.9531$ for single compression.
For Top-\{2,3\} predictions, the performance never falls below $0.9496$.
The difference to the remaining two backbones is the largest for the Top-1 accuracy.
Here, the scores of GamperCNN are on average lower by 4.92 \gls{pp}, while Götz is on average even 6.91 \gls{pp} behind Gamper$^\star$.
All in all, the experiment thus shows EnvId's good generalization ability to longer compression chains, especially with the proposed Gamper$^\star$ backbone.}

\subsubsection{Unseen Compression Codecs} Fast progress in audio compression research might lead to the case that EnvId must perform inference on compression codecs that it has not been trained for.

\edited{We thus evaluate EnvId on signals compressed by two codecs that
were not seen during training, namely Vorbis as an established analytic compressor and EnCodec~\cite{defossez2022highfi} from Meta Research\nolinebreak\textsuperscript{\copyright} as a relevant neural network-based compressor.
The data generation pipeline is configured accordingly to produce in total $6$ MIT test set variants, one for each codec with low, middle and high quality settings.
No noise degradation is applied to the samples.}

Fig.~\ref{subfig:vorbis} shows the result for Vorbis, and Fig.~\ref{subfig:encodec} shows the results for EnCodec.
The accuracy on Vorbis compression is very close to $1$ for the best performing Gamper$^\star$, which is
understandable, as it targets mid to high quality output where the lowest available bitrate is still as high as $16$\,kbps.
In line with the results on multi-compression, GamperCNN and Götz perform second and third best, and the difference to Gamper$^\star$ is most notable for the Top-1 accuracy, where, on average, GamperCNN and Götz perform lower than Gamper$^\star$ by 4.92 \gls{pp} and 5.89 \gls{pp}. 

\edited{In contrast to Vorbis, EnCodec has a higher impact on the performance.
Overall, the Gamper$^\star$ backbone has a slight advantage over the remaining two backbones on the high ($24$\,kbps) quality setting, and performs second best but comparably to GamperCNN for the mid ($6$\,kbps) and low ($1.5$\,kbps) quality setting.
More precisely, the Top-1 accuracies of Gamper$^\star$ on the high ($24$\,kbps) and mid ($6$\,kbps) quality settings are $0.8951$ and
$0.8028$, but it drops to $0.5478$ for the low ($1.5$\,kbps) quality setting. 
Also, GamperCNN only reaches slightly better Top-1 scores of $0.8131$ and $0.5713$ for mid and low qualities.
The low accuracy on such an extremely low bitrate is not surprising: the lowest available bitrate in the training set are
$4.75$\,kbps with AMR-NB, hence EnvId has to simultaneously deal with unseen
compression artifacts at an unseen low bitrate.}

\edited{In total, we report that EnvId is quite robust to unseen compression codecs, also at low bitrates of $24$\,kbps or even $6$\,kbps.
Only EnCodec's strongest compression of $1.5$\,kbps considerably degrades the identification accuracy.  }

\subsubsection{Unseen Noise Impact}
In practice, the background noise may considerably vary in the training and test data distribution. To evaluate EnvId's generalization ability towards real environmental background noise,
a selection of $4$ publicly available noise sources
is used that cover a passing train\footnote{\url{https://freesound.org/people/theplax/sounds/615849/}}, falling rain\footnote{{\url{https://freesound.org/people/straget/sounds/531947/}}},
and sounds of a crowded airport\footnote{\url{https://freesound.org/people/arnaud\%20coutancier/sounds/424362/}} and exhibition hall\footnote{\url{https://freesound.org/people/BockelSound/sounds/487600/}}.
\edited{For the evaluation, $3$ MIT test set versions with low (\gls{snr}~$=50$\,dB), medium (\gls{snr}~$=25$\,dB) and high (\gls{snr}~$=0$\,dB) noise degradation a are generated.
For each set version, one of the $4$ background noise patterns is randomly selected per sample and added with the corresponding fixed test \gls{snr}. }

\edited{The evaluation result is shown in Fig.~\ref{subfig:noise}.
In total, the Gamper$^\star$ backbone again performs best, followed by GamperCNN and Götz. 
Solely for strong noise impact with \gls{snr}~$= 0$\,dB, GamperCNN$^\star$ performs approximately equally well to Gamper$^\star$.
Low (\gls{snr}~$= 50$\,dB) and medium (\gls{snr}~$= 25$\,dB) noise levels have little influence on EnvId's Top-1 accuracy which is especially close to $1$ for Gamper$^\star$.
However, strongly degraded signals (\gls{snr}~$= 0$\,dB) cause the accuracy to fall to $0.7648$, $0.8697$ and $0.9134$ for the Top-1, Top-2 and Top-3 predictions, respectively. 
Nevertheless, the results are remarkable when considering that the noise pattern is unseen and the distortion of the signal is extremely high.}

\subsection{\edited{Ablation on the Number of Samples per Prototypes}}
\label{subsec:k_ablation}

EnvId requires a support set of $K$ reference samples for prototype computation per candidate recording environment.
There are no strict constraints on the number of reference samples. 
From a theoretical perspective, one might expect that ``more data is better''.
\edited{However, in practice, meta-learning approaches like Prototypical Networks tend to overfit on the $K$ value used in training~\cite{Cao2020A}.
Thus, $K$ is typically chosen to be identical in training and testing as done in our experiments (\cf Sec.~\ref{subsec:experimental_setup}).}
Still, the question of a lower limit on the number of required samples is relevant for our scenario.
After all, the collection of reference samples from locations in a forensic case is arguably constrained by the available resources for data acquisition.
Therefore, we here evaluate the impact of testing with $K < 15$ using our EnvId framework trained with $K=15$ samples (\cf Sec.~\ref{subsec:experimental_setup}).

Fig.~\ref{fig:kshot_plot} shows the accuracy of EnvId for $K$ varying between $1$ and $15$, using the best performing feature extractor Gamper$^\star$ on the $4$ test sets from Sec.~\ref{subsubsec:dataset_benchmark}. 
For our scenario, the results do not indicate a severe impact of a train/test mismatch of $K$ for most test sets.
In detail, for ACE and AIR+REV sets, already $K=1$ provides a good prototype and locations of MIT can be stably identified with $K=2$ reference samples. 
\edited{Only OPENAIR locations
	require at least $K = 10$ samples to yield stable prototypes.
	OPENAIR's mostly large and strongly reverberant environments deviate the most from the training set distribution (\cf Sec~\ref{subsubsec:gen_ability}).
	Thus, we hypothesize that EnvId's embedding space is less fit for representing these types of locations, such that a higher $K$ is needed to form stable prototypes.
	Analogously to Sec~\ref{subsubsec:gen_ability}, these findings emphasize the importance of analyzing training data for weakly covered environmental characteristics to prevent such pitfalls.}

\begin{figure}[t]
	\resizebox{0.49\textwidth}{!}{\includesvg{plots/k_shot.svg}}
	\caption{Location identification accuracy w.r.t. the $K$-shot parameter for inference on the test sets from Sec.~\ref{subsubsec:dataset_benchmark}. The out-of-distribution OPENAIR environments require the most $K$ reference samples.}
	\label{fig:kshot_plot}
\end{figure}

\begin{figure*}[t]
	\newcommand{\imgscalew}{0.34}
	\newcommand{\imgscaleh}{0.34}
	\hspace{-0.6cm}
	\subfloat[Category \emph{corridor}\label{subfig:corridor}]{\includesvg[width=0.37\textwidth]{plots/hall-big-proportions}}\hspace{-0.5cm}
	\subfloat[Category \emph{rectangle}\label{subfig:rectangle}]{\includesvg[width=0.37\textwidth]{plots/rectangle-big-proportions}}\hspace{-0.5cm}
	\subfloat[Category \emph{square}\label{subfig:square}]{\includesvg[width=0.37\textwidth]{plots/square-big}}

	\caption{\edited{Mean accuracy for specific microphone positions on a uniformly sampled $5 \times 5$ grid for the categories \emph{corridor}, \emph{rectangle} and \emph{square}. Environments for audio samples recorded closer to boundaries are on average easier to identify. The total accuracy thus decreases for wider room widths from left to right.}}
	\label{fig:gird_map}
\end{figure*}

\subsection{Recording Position Mismatch}
\label{subsec:synth_stuff}
The actual position of the microphone in the room is a subtle, yet quite
impactful issue that has been barely addressed in related works (\cf Sec.~\ref{subsec:limitations_related_works}).

If reference audio samples from candidate environments are collected during the investigation process, it must be assumed that the recording positions in the support set is at least somewhat different from the query signal.
Therefore, in this experiment, we provide a benchmark and data set with synthetic data of enclosed spaces of varying recording positions in differently shaped rooms.

\edited{For the evaluation, we fine-tune EnvId with the best performing Gamper$^\star$ feature extractor from Sec.~\ref{subsec:extractor_benchmark_exp} on a synthetic data set with annotated recording positions.
The training hyperparameters remain unchanged for the fine-tuning process (\cf Sec.~\ref{subsec:experimental_setup}).}

\edited{The experiment is split into four parts. 
In Sec.~\ref{subsubsec:room_simulation}, the method for simulating \glspl{air} from synthetic rooms is described and Sec.~\ref{subsubsec:synth_data} summarizes the resulting training and test data sets of reverberant speech.
The results on the influence of the microphone position is presented in Sec.~\ref{subsubsec:micpos_eval}.
Sec.~\ref{subsubsec:synth_robustness} further investigates the impact of noise and compression signal degradation, and Sec.~\ref{subsubsec:synth_discussion} discusses the results.}

\subsubsection{Simulation of Synthetic AIRs}
\label{subsubsec:room_simulation}
To our knowledge, there is no data set that provides a grid of real \gls{air} measurements with known positions for a large number of recording locations.
Hence, we simulate enclosed spaces with pyroomacoustics~\cite{scheibler2018pyroomacoustics}.
We consider rooms within a length interval of $l_r \in [1, 50]$\,m and height $h_r \in [2, 5]$\,m  in steps of $10$\,cm.
Three common room shapes are \emph{corridor}, \emph{rectangle}, and \emph{square} rooms, where we define the fraction $f$ of width over length as $f \in [0.1, 0.3]$ for corridor,  $f \in [0.4, 0.7]$ for rectangle, and $f \in [0.8, 1]$ for square.

Different floor and wall characteristics are simulated with an absorption coefficient $c_a \in [0.1, 0.8]$. 
\gls{air} measurements are sampled from a equidistant $5\times5$ grid per room to uniformly cover potential microphone positions. 
The distance between the outer edges of the grid to the corresponding nearest wall is set to $30$\,cm.
For this first study, we focus on the practically relevant scenario of some speaker during a telephone conversation or the recording of voice messages.
The relative distance between the microphone and the speaker is thus set to a fixed size of $10$\,cm, and the microphone is always positioned $1.7$\,m above the ground floor.  
The speaker orientation is always set towards the room center and the microphone is set inversely.
This preserves the comparability of different grid positions.

\subsubsection{Training and Test Data}
\label{subsubsec:synth_data}
 Our final training/validation includes samples from $45$/$15$ rooms for each of the $3$ shape categories uniformly sampled from the whole volume range, which leads to $135$/$45$ rooms per set. 
For each room, absorption coefficients are randomly sampled. 
\gls{air} measurements are simulated for each of the $5 \times 5$ grid positions, yielding a total of $3375$/$1125$ training/validation \glspl{air}.
Each \gls{air} is convolved with each of $25$/$19$ anechoic speech samples from a subset of the ACE training and validation pool (Tab.~\ref{tab:data_pools}). 
To approximate real world scenarios, the data generation pipeline is configured to randomly add white noise within a \gls{snr} range of $\alpha \in [-10, 50]$  and perform arbitrarily strong MP3, AMR-NB or GSM single compression for each sample.
This degradation configuration is analogous to the training set from the previous experiments (\cf Sec.~\ref{subsubsec:dataset_benchmark}).
The final sets consist of about $84$k/$21$k training/validation samples.	

The test set is constructed with a volume range of $V \in [10, 3750]$\,m$^3$, as $3750$\,m$^3$ is the maximum possible volume for a \emph{corridor} room in our setting. 
We sample one room per shape category for $10$ volumes uniformly distributed over the full range to yield $30$ test rooms in total.
The absorption coefficient is fixed to $c_a = 0.1$ for medium absorption~\cite{scheibler2018pyroomacoustics}.
For this experiment, the fixed absorption coefficient in the test set enables an isolated analysis of the impact of changing recording positions.
Per room, \glspl{air} measured from the $5 \times 5$ grid positions are convolved each with a subset of $16$ samples from the TSP test pool (Tab.~\ref{tab:data_pools}), yielding $12$k samples in total. 
The test set does not include noise or compression degradation, whose influence is analyzed individually in Sec.~\ref{subsubsec:synth_robustness}.

\subsubsection{\edited{Recording Position Impact}}
\label{subsubsec:micpos_eval}
For the evaluation, the samples from one position (queries) are tested against prototypes that are calculated randomly from reference samples of all other recording positions.
When averaging over all positions in all rooms per room category, then
\emph{corridor} rooms are easiest to identify with an average accuracy of
$0.7715$, followed by \emph{rectangle} and \emph{square} rooms with average
accuracies of $0.7648$ and $0.7608$.

The results for each individual position are visualized in Fig.~\ref{fig:gird_map} for the three room shape
categories \emph{corridor} (left), \emph{rectangle} (middle), and \emph{square}
(right).
We additionally provide numerical values for each position in Tab.\,A.4 in the supplemental material.\textsuperscript{2}
Overall, the accuracy tends to increase at locations that are closer to walls.
For example, the average center accuracies are $0.6375$, $0.6875$ and $0.7063$
for \emph{corridor}, \emph{rectangle} and \emph{square} rooms, but average
accuracies at corners are $0.8453$, $0.7750$ and $0.7328$.
\edited{For \emph{rectangle} and \emph{square} rooms it is further notable that the most difficult line of positions is the one along the middle column, on average $0.7038$ and $0.7113$. For \emph{corridor} rooms, the middle row is the most difficult grid line with $0.6800$.
Overall, the difference in accuracy between inner and outer positions is higher for the narrow \emph{corridor} spaces than for the more uniform \emph{rectangle} or \emph{square} rooms.
The larger variation is reflected by the standard deviation over all positions on the grid, which is $0.0748$ for the category \emph{corridor}, but only $0.0419$ and $0.0438$ for the category \emph{rectangle} and \emph{square}. 
}
These findings suggest that early reflections in the \gls{air} for recordings close to walls lead to more characteristic features, which makes environment identification easier.
This is in line with the observation that the overall accuracies are highest on \emph{corridor} rooms, where the distances to the side walls are relatively short.

\subsubsection{\edited{Influence of Noise and Compression}}
\label{subsubsec:synth_robustness}
We also investigate the impact of compression and noise degradation on the identification scores.
\edited{Therefore, two further versions of our test set of synthetic rooms from Sec.~\ref{subsubsec:synth_data} are generated. For the compression test set, the pipeline is again configured to randomly compress each sample once with MP3, AMR-NB or GSM and random compression quality settings, and for the noise test set, additive white noise of random \gls{snr} between $-10$\,dB and $50$\,dB is chosen for each sample. }

For compression, the identification scores for the three room categories, \ie, the averaged accuracy over all recording positions drops to $0.6458$, $0.6163$ and $0.6340$ for \emph{corridor}, \emph{rectangle} and \emph{square} rooms.
This corresponds to an average decrease by $13.67$ \gls{pp}.
Noise has an even stronger impact and decreases the values down to $0.5298$, $0.4898$ and $0.5118$, \ie, a decrease by $25.52$ \gls{pp}.

\subsubsection{\edited{Discussion}}
\label{subsubsec:synth_discussion}
\edited{The results indicate that recording position is indeed an important factor for environment identification that should be considered in future research and in practice.
Especially speech that is recorded near walls is easier to identify due to nearby signal reflections.
On average, this leads to better results on elongated rooms like corridors than on square rooms of the same volume.
In future work, these findings remain to be more closely analyzed by acquiring a corpus of real-world \gls{air} measurements from annotated positions.}

\subsection{Blind Environment Parameter Estimation}
\label{subsec:param_est}
\edited{EnvId blindly regresses environmental parameters from the query's embedding vector (\cf Sec.~\ref{subsec:envid_methods}).
Such parameter estimates can act as an additional hint to the few-shot location predictions.
Additionally, the regression can still provide valuable cues for forensic cases, where no set of candidate locations is available.
}
In our evaluation, we demonstrate the regression capability at the example of the room volume $V$ and the RT$_{60}$ parameter, which are popular
quantities in related work (\cf Sec.~\ref{subsec:existing_methods}).

\edited{We use the same training setup as for the previous experiment, \ie, train the best EnvId configuration on our set of synthetic enclosed spaces (\cf Sec.~\ref{subsubsec:synth_data}).
For the synthetic data, both volume and RT$_{60}$ labels are known, such that EnvId's regression head can be jointly trained with the few-shot task (\cf Sec.~\ref{subsubsec:envid_training}).}

\subsubsection{\edited{Comparison to Related Metric-Learning Approach}} 
\label{subsubsec:synth_baseline}
We first compare EnvId's regression capability to the recently proposed parameter estimator by Götz~\etal~\cite{gotz2023contrastive} who also apply representation learning techniques.

In detail, they use contrastive learning with an upstream encoder network to obtain embeddings for room volume and RT$_{60}$. 
These embeddings are fed to a downstream network to
classify the room volume as small or large, and to regress the RT$_{60}$.  
With the help of the authors, we recreate the data set from the publication according
to their protocol. The training set consists of simulated rooms of uniformly
distributed volumes $V \in [27 ,500]$m$^{3}$ and randomly sampled absorption
coefficients. The test set consists of $1000$ \glspl{air}, \ie, $10$
\glspl{air} from $100$ simulated rooms as described in their
work~\cite{gotz2023contrastive} .

When running this experiment, 
EnvId achieves a \gls{rmse} of $0.2129$ for RT$_{60}$ and a remarkable volume
classification accuracy of $0.9998$. 
Both values outperform the approach of
G{\"o}tz~\etal, who report a minimum \gls{rmse} of $0.2146$ for RT$_{60}$ and a
maximum volume classification accuracy of $0.7422$. 

\edited{To analyze the capabilities of EnvId more in detail, we additionally evaluate the model on our synthetic rooms with larger volume range  $V \in [10, 3750]$m$^3$.
To this end, we construct $3$ test sets.
One version with clean samples is generated like the test set for the previous experiment (\cf Sec.~\ref{subsubsec:synth_data}), however with randomly sampled absorption coefficients $c_a$ between $0.1$ and $0.8$ for all rooms.
We also generate a noisy version and a version with single-compressed signals. The results are presented in the following two sections.
}
\begin{table}[t]
	\centering
	\caption{Results for the regression of RT$_{60}$ on our $3$ synthetic test sets of clean, single compressed and noisy signals.  The \gls{rmse} and mean target RT$_{60}$ are reported grouped by the room categories \emph{corridor}, \emph{rectangle}, and \emph{square}.}
	\label{tab:room_params_ours}
	\begin{tabular}{lcccc}
		
		\toprule[1pt]
		\multirowcell{2}{Room \\ Category} & \multicolumn{3}{c}{RMSE} & \multicolumn{1}{c}{\multirowcell{2}{Mean \\ Targets}} \\
		\cmidrule{2-4}
		& Clean & Compr. & Noise &    \\
		
		\midrule[1pt]

		Corridor  & 0.2640 & 0.2723 & 0.2748 & 0.6497 \\
		Rectangle & 0.1692 & 0.1921 & 0.2157 & 0.4930 \\
		Square    & 0.1553 & 0.1777 & 0.2117 & 0.4125 \\

		\bottomrule[1pt]   
	\end{tabular}
	
\end{table}

\begin{figure}[t]
	\resizebox{0.49\textwidth}{!}{\includesvg{plots/rmse_per_absorption.svg}}
	\caption{\edited{\gls{rmse} results for RT$_{60}$ grouped by absorption coefficient $c_a \in [0.1, 0.8]$ for clean, noisy and single compressed test samples. }}\label{fig:rmse_per_absorption}
\end{figure}

\subsubsection{\edited{Analysis of RT$_{60}$ Estimations}}
\label{subsubsec:synth_rt60}

We first evaluate the influence of room shape on the estimation of the RT$_{60}$ parameter. 
The results are shown in Tab.~\ref{tab:room_params_ours}. 
Both the \gls{rmse} and RT$_{60}$ mean values of the targets are reported grouped by shape category.
As is reflected by the mean target RT$_{60}$, the  RT$_{60}$ naturally increases for more elongated rooms in our set. 
Equally, the \gls{rmse} increases, such that \emph{corridor} rooms exhibit higher \gls{rmse} scores than \emph{rectangle} and especially \emph{square} rooms. 
Compression and added noise slightly reduce the performance. The average \gls{rmse} increases by $1.79$ \gls{pp} for compression and by $3.79$ \gls{pp} for noise. 
These finding are in line with the results in the previous experiment (Sec.~\ref{subsec:synth_stuff}).

\edited{Additionally to room shape and size, the absorption coefficient $c_a$ has a large impact on the \gls{rmse} predictions. 
The absorption characteristic for a specific room geometry strongly influences the energy decay of some emitted signal and thus the resulting RT$_{60}$ value.
Fig.~\ref{fig:rmse_per_absorption} shows the average \gls{rmse} scores grouped by $c_a$ and signal degradation in the test data.
Note that for this experiment, we conduct in total   $8 \cdot 3 = 24$ runs, \ie, one run on each of $8$ variants of the clean, noise and single compression test data sets with fixed $c_a \in [0.1, 0.2, \cdots, 0.8]$.
Evidently, the \gls{rmse} decreases with increasing absorption, \ie, decreasing RT$_{60}$ labels which complements the results in the last column of Tab.~\ref{tab:room_params_ours}.
Also, in line with previous experiments, single compression and noise degradation increase the prediction errors. }

\begin{table}[]
	\centering	
	\caption{Volume classification results on our $3$ synthetic test sets with clean, single compressed and noisy samples. We report scores for classifying rooms in $2$, $3$, $5$, or $10$ size categories.}\label{tab:volume_classification}
	\begin{tabular}{lcccc}
		\toprule[1pt]
		\multirowcell{2}{Signal \\ Degradation}&\multicolumn{4}{c}{Accuracy}           \\ 
		\cmidrule{2-5}
		& 2-class & 3-class & 5-class  & 10-class \\ \midrule[1pt]
		Clean & 0.9706 & 0.9519        &  0.8566       &  0.7808       			\\
		Compression &0.8872&0.8410&0.7045&0.6593 \\
		Noise & 0.8298&0.7930&0.6605&0.5999\\
		\bottomrule[1pt]
	\end{tabular}
	
\end{table}

\subsubsection{\edited{Fine-Grained Volume Classification}}
\label{subsubsec:synth_volume}
To provide an intuition about the quality of the volume predictions, we extend
the evaluation approach by Götz~\etal~\cite{gotz2023contrastive}.
As previously discussed, they use the volume estimate to classify room sizes as small or large (\cf Sec.~\ref{subsubsec:synth_baseline}).
\edited{In this experiment, we generalize this approach beyond a binary task and evenly divide the volume range of our synthetic rooms in $n \in \{2, 3, 5, 10\}$ classes to test increasingly fine-grained versions of the problem.
Equivalently to Götz~\etal~\cite{gotz2023contrastive}, the classification is done via binning the volume estimates, \ie, classifying the volume as belonging to one of the $n$ size categories. 
}

Tab.~\ref{tab:volume_classification} shows the results on our $3$ test set versions with clean, noisy and single compressed samples.
The two class task, \ie, $n = 2$ is solved with a high accuracy of $0.9706$, also for our data set.
Single compression decreases the accuracy to $0.8872$, and added noise to $0.8298$.
Naturally, the accuracy decreases for increasing numbers of size categories.
Nevertheless, for the finest problem granularity of $n = 10$ classes, EnvId still
achieves accuracies of $0.7808$, $0.6593$ and $0.5999$ for clean, compressed
and noisy inputs.  

\subsubsection{\edited{Discussion}}
\label{subsubsec:synth_reg_discussion}
Overall, our findings show that EnvId is able to blindly extract relevant cues about recording location characteristics from the queries' embedded representations, also under signal impacting factors like noise or compression. 
Additionally, our experiments demonstrate that optimizing the regressor in EnvId as a side-task to few-shot identification can outperform a state-of-the-art approach for parameter estimation only.

\section{Conclusion}
\label{sec:conclusion}
\gls{dl} tools are increasingly moving into the focus of police authorities to
support criminal investigations.  However, it remains an open challenge to
develop tools that are flexible and robust enough to meet forensic
requirements.  
In particular, the identification of recording environments from
single audio samples is difficult due to the many practical challenges in
this seemingly straightforward task.

In this work, we propose the end-to-end trainable EnvId framework as a step towards practically applicable recording environment identification. 
The proposed framework performs few-shot classification to identify case-specific environments not seen in training and supports environmental parameter estimation for recording location characterization.

Furthermore, EnvId addresses open-set identification and identification of data with unknown signal degradations, which is equally motivated by the challenges in forensic police work.
We also explore the so-far neglected, but practically very important issue of a mismatch in recording positions within a location. Here,
our experiments quantitatively show that it is notably easier to identify a
room from mismatched recording position when they were done closer to the
walls. 

\edited{We see EnvId as a cornerstone for future work on these challenges. 
We aim to further explore the task of forensic recording location identification and improve our framework's overall performance.
For example, the efficiency on real-world reverberant speech samples from annotated locations is yet to be explored.
Up to now, this is hindered by the lack of sufficiently large available data sets.
Additionally, a protocol on how to collect an optimal support set of reference samples from recording locations for practice has to be established. 
After all, recording settings like the microphone position can influence the few-shot performance, as suggested by our experiments.}
\FloatBarrier
\bibliographystyle{plain}
\bibliography{mybib,mybib2}


\end{document}

%% file: svg-inkscape/pipeline_svg-tex.pdf_tex
\begingroup%
  \makeatletter%
  \providecommand\color[2][]{%
    \errmessage{(Inkscape) Color is used for the text in Inkscape, but the package 'color.sty' is not loaded}%
    \renewcommand\color[2][]{}%
  }%
  \providecommand\transparent[1]{%
    \errmessage{(Inkscape) Transparency is used (non-zero) for the text in Inkscape, but the package 'transparent.sty' is not loaded}%
    \renewcommand\transparent[1]{}%
  }%
  \providecommand\rotatebox[2]{#2}%
  \newcommand*\fsize{\dimexpr\f@size pt\relax}%
  \newcommand*\lineheight[1]{\fontsize{\fsize}{#1\fsize}\selectfont}%
  \ifx\svgwidth\undefined%
    \setlength{\unitlength}{657.75bp}%
    \ifx\svgscale\undefined%
      \relax%
    \else%
      \setlength{\unitlength}{\unitlength * \real{\svgscale}}%
    \fi%
  \else%
    \setlength{\unitlength}{\svgwidth}%
  \fi%
  \global\let\svgwidth\undefined%
  \global\let\svgscale\undefined%
  \makeatother%
  \begin{picture}(1,0.24048175)%
    \lineheight{1}%
    \setlength\tabcolsep{0pt}%
    \put(0,0){\includegraphics[width=\unitlength,page=1]{svg-inkscape/pipeline_svg-tex.pdf}}%
    \put(0.12200684,0.07172177){\color[rgb]{0,0,0}\makebox(0,0)[t]{\lineheight{1.25}\smash{\begin{tabular}[t]{c}$a(t)$\end{tabular}}}}%
    \put(0,0){\includegraphics[width=\unitlength,page=2]{svg-inkscape/pipeline_svg-tex.pdf}}%
    \put(0.15963512,0.12873432){\color[rgb]{0,0,0}\makebox(0,0)[t]{\lineheight{1.25}\smash{\begin{tabular}[t]{c}AIRs\end{tabular}}}}%
    \put(0,0){\includegraphics[width=\unitlength,page=3]{svg-inkscape/pipeline_svg-tex.pdf}}%
    \put(0.21322691,0.15039908){\color[rgb]{0,0,0}\makebox(0,0)[t]{\lineheight{1.25}\smash{\begin{tabular}[t]{c}$r(t)$\end{tabular}}}}%
    \put(0,0){\includegraphics[width=\unitlength,page=4]{svg-inkscape/pipeline_svg-tex.pdf}}%
    \put(0.44127708,0.15039908){\color[rgb]{0,0,0}\makebox(0,0)[t]{\lineheight{1.25}\smash{\begin{tabular}[t]{c}$n(t)$\end{tabular}}}}%
    \put(0,0){\includegraphics[width=\unitlength,page=5]{svg-inkscape/pipeline_svg-tex.pdf}}%
    \put(0.35005701,0.07172177){\color[rgb]{0,0,0}\makebox(0,0)[t]{\lineheight{1.25}\smash{\begin{tabular}[t]{c}$s(t)$\end{tabular}}}}%
    \put(0,0){\includegraphics[width=\unitlength,page=6]{svg-inkscape/pipeline_svg-tex.pdf}}%
    \put(0.66932725,0.15039908){\color[rgb]{0,0,0}\makebox(0,0)[t]{\lineheight{1.25}\smash{\begin{tabular}[t]{c}$f(t)$\end{tabular}}}}%
    \put(0,0){\includegraphics[width=\unitlength,page=7]{svg-inkscape/pipeline_svg-tex.pdf}}%
    \put(0.57810718,0.07172177){\color[rgb]{0,0,0}\makebox(0,0)[t]{\lineheight{1.25}\smash{\begin{tabular}[t]{c}$\hat{s}(t)$\end{tabular}}}}%
    \put(0,0){\includegraphics[width=\unitlength,page=8]{svg-inkscape/pipeline_svg-tex.pdf}}%
    \put(0.80615735,0.07172177){\color[rgb]{0,0,0}\makebox(0,0)[t]{\lineheight{1.25}\smash{\begin{tabular}[t]{c}$\tilde{s}(t)$\end{tabular}}}}%
    \put(0,0){\includegraphics[width=\unitlength,page=9]{svg-inkscape/pipeline_svg-tex.pdf}}%
    \put(0.49828962,0.15039908){\color[rgb]{0,0,0}\makebox(0,0)[t]{\lineheight{1.25}\smash{\begin{tabular}[t]{c}$\alpha$\end{tabular}}}}%
    \put(0,0){\includegraphics[width=\unitlength,page=10]{svg-inkscape/pipeline_svg-tex.pdf}}%
    \put(0.51311288,0.18688711){\color[rgb]{0,0,0}\makebox(0,0)[t]{\lineheight{1.25}\smash{\begin{tabular}[t]{c}SNRs\end{tabular}}}}%
    \put(0,0){\includegraphics[width=\unitlength,page=11]{svg-inkscape/pipeline_svg-tex.pdf}}%
    \put(0.72633979,0.15039908){\color[rgb]{0,0,0}\makebox(0,0)[t]{\lineheight{1.25}\smash{\begin{tabular}[t]{c}$\mathcal{C}$\end{tabular}}}}%
    \put(0,0){\includegraphics[width=\unitlength,page=12]{svg-inkscape/pipeline_svg-tex.pdf}}%
    \put(0.74800456,0.036374){\color[rgb]{0,0,0}\makebox(0,0)[t]{\lineheight{1.25}\smash{\begin{tabular}[t]{c}$N\times$\end{tabular}}}}%
    \put(0,0){\includegraphics[width=\unitlength,page=13]{svg-inkscape/pipeline_svg-tex.pdf}}%
    \put(0.03762828,0.06031927){\color[rgb]{0,0,0}\makebox(0,0)[t]{\lineheight{1.25}\smash{\begin{tabular}[t]{c}Anechoic\end{tabular}}}}%
    \put(0,0){\includegraphics[width=\unitlength,page=14]{svg-inkscape/pipeline_svg-tex.pdf}}%
    \put(0.03762828,0.0432155){\color[rgb]{0,0,0}\makebox(0,0)[t]{\lineheight{1.25}\smash{\begin{tabular}[t]{c}Speech\end{tabular}}}}%
    \put(0,0){\includegraphics[width=\unitlength,page=15]{svg-inkscape/pipeline_svg-tex.pdf}}%
    \put(0.38540479,0.13899658){\color[rgb]{0,0,0}\makebox(0,0)[t]{\lineheight{1.25}\smash{\begin{tabular}[t]{c}Noise\end{tabular}}}}%
    \put(0,0){\includegraphics[width=\unitlength,page=16]{svg-inkscape/pipeline_svg-tex.pdf}}%
    \put(0.38540479,0.12189281){\color[rgb]{0,0,0}\makebox(0,0)[t]{\lineheight{1.25}\smash{\begin{tabular}[t]{c}Samples\end{tabular}}}}%
    \put(0,0){\includegraphics[width=\unitlength,page=17]{svg-inkscape/pipeline_svg-tex.pdf}}%
    \put(0.61573546,0.13671607){\color[rgb]{0,0,0}\makebox(0,0)[t]{\lineheight{1.25}\smash{\begin{tabular}[t]{c}Compr.\end{tabular}}}}%
    \put(0,0){\includegraphics[width=\unitlength,page=18]{svg-inkscape/pipeline_svg-tex.pdf}}%
    \put(0.61573546,0.11961231){\color[rgb]{0,0,0}\makebox(0,0)[t]{\lineheight{1.25}\smash{\begin{tabular}[t]{c}Formats\end{tabular}}}}%
    \put(0,0){\includegraphics[width=\unitlength,page=19]{svg-inkscape/pipeline_svg-tex.pdf}}%
    \put(0.74116306,0.19372862){\color[rgb]{0,0,0}\makebox(0,0)[t]{\lineheight{1.25}\smash{\begin{tabular}[t]{c}Compr.\end{tabular}}}}%
    \put(0,0){\includegraphics[width=\unitlength,page=20]{svg-inkscape/pipeline_svg-tex.pdf}}%
    \put(0.74116306,0.17662485){\color[rgb]{0,0,0}\makebox(0,0)[t]{\lineheight{1.25}\smash{\begin{tabular}[t]{c}Strength\end{tabular}}}}%
    \put(0,0){\includegraphics[width=\unitlength,page=21]{svg-inkscape/pipeline_svg-tex.pdf}}%
    \put(0.23603193,0.06944127){\color[rgb]{0,0,0}\makebox(0,0)[t]{\lineheight{1.25}\smash{\begin{tabular}[t]{c}\textbf{(a)} Signal Reverb.\end{tabular}}}}%
    \put(0,0){\includegraphics[width=\unitlength,page=22]{svg-inkscape/pipeline_svg-tex.pdf}}%
    \put(0.23603193,0.05119726){\color[rgb]{0,0,0}\makebox(0,0)[t]{\lineheight{1.25}\smash{\begin{tabular}[t]{c}$r(t) * a(t)$\end{tabular}}}}%
    \put(0,0){\includegraphics[width=\unitlength,page=23]{svg-inkscape/pipeline_svg-tex.pdf}}%
    \put(0.4640821,0.06944127){\color[rgb]{0,0,0}\makebox(0,0)[t]{\lineheight{1.25}\smash{\begin{tabular}[t]{c}\textbf{(b)} Noise Impact\end{tabular}}}}%
    \put(0,0){\includegraphics[width=\unitlength,page=24]{svg-inkscape/pipeline_svg-tex.pdf}}%
    \put(0.4640821,0.05119726){\color[rgb]{0,0,0}\makebox(0,0)[t]{\lineheight{1.25}\smash{\begin{tabular}[t]{c}$s(t) + \alpha \cdot n(t)$\end{tabular}}}}%
    \put(0,0){\includegraphics[width=\unitlength,page=25]{svg-inkscape/pipeline_svg-tex.pdf}}%
    \put(0.69213227,0.06944127){\color[rgb]{0,0,0}\makebox(0,0)[t]{\lineheight{1.25}\smash{\begin{tabular}[t]{c}\textbf{(c)} Compr. Impact\end{tabular}}}}%
    \put(0,0){\includegraphics[width=\unitlength,page=26]{svg-inkscape/pipeline_svg-tex.pdf}}%
    \put(0.69213227,0.05119726){\color[rgb]{0,0,0}\makebox(0,0)[t]{\lineheight{1.25}\smash{\begin{tabular}[t]{c}$f^{\mathcal{C}}_n(\hat{s}(t))$\end{tabular}}}}%
    \put(0,0){\includegraphics[width=\unitlength,page=27]{svg-inkscape/pipeline_svg-tex.pdf}}%
  \end{picture}%
\endgroup%